\documentclass[aps,preprint,prd,epsfig,nofootinbib]{revtex4}
\pdfoutput=1
\usepackage{amssymb}
\usepackage{amsmath}
\usepackage{color}
\usepackage{fancyhdr}
\newcommand{\be}{\begin{equation}}
\newcommand{\ee}{\end{equation}}
\newcommand{\bea}{\begin{eqnarray}}
\newcommand{\eea}{\end{eqnarray}}
\newcommand{\bes}{\begin{subequations}}
\newcommand{\ees}{\end{subequations}}

\newcommand{\bc}{\begin{center}}
\newcommand{\ec}{\end{center}}
\usepackage{amsmath, amssymb} 
\usepackage{slashed}
\usepackage{graphicx}
\usepackage{subcaption}
\begin{document}

\title{On the Higgs spectra of the  3-3-1 model with the sextet of scalars engendering the type II seesaw mechanism }

\author{Jo\~ao Paulo Pinheiro$^{a}$, C. A. de S. Pires$^{b}$}
\affiliation{{$^a$ Departament de Física Quàntica i Astrofísica and Institut de Ciències del Cosmos, Universitat
de Barcelona, Diagonal 647, E-08028 Barcelona, Spain},  \\
{$^b$  Departamento de Física, Universidade Federal da Paraíba, Caixa Postal 5008, 58051-970, Jo\~ao Pessoa, PB, Brazil}, \\
}

\date{\today}

\begin{abstract}

In the 3-3-1 model with right-handed neutrinos (331RHNs), three triplets of scalars engender the correct sequence of symmetry breaking, $SU(3)_C \times SU(3)_L \times U(1)_X \rightarrow SU(3)_C \times SU(2)_L \times U(1)_Y \rightarrow SU(3)_C \times U(1)_{\mathrm{EM}}$, generating mass for all  fermions, except neutrinos. Tiny neutrino  masses  may be achieved by implementing the type II seesaw mechanism into the model, requiring the addition of  one sextet of scalars to the original scalar content. As consequence, it emerges a very complex scalar sector, involving terms that violate lepton number explicitly, too. The main obstacle to the development of the phenomenology of such scenario is the knowledge  of its spectrum of scalars since, now, there are 15 massive scalar particles on it.  The proposal of this work is to do an exhaustive analysis of such scalar sector with lepton number being explicitly violated at low, electroweak and high energy scales by means of trilinear terms in the potential. The first case can be addressed analytically and, as a nice result, we have observed that the scalar content of such case is split into two categories: One belonging to the 331 energy scale and the other belonging to the EWSB energy scale, with the last  recovering the well known THDM + triplet. For the  other cases, the scalar sector can be  addressed only numerically. Hence, we proposed a very general approach for the numerical study of the potential, avoiding simplifications that can make us reach conclusions without foundation. We show that, in the case of lepton number being explicitly violated at electroweak scale, it is possible to recover the same physics of the THDM + triplet, as the previous case. Among all the possibilities, we call the attention to one special case  which generates the 3HDM+triplet scenario. For the last case, when lepton number is violated at high energy scale, the sextet become very massive and  decouples from the original scalar content of the 3-3-1 model. We also discuss, when possible, the phenomenological aspects of each case.

\end{abstract}

\maketitle
\section{Introduction}
The type II seesaw\cite{Ma:1998dx}  is an elegant and  versatile mechanism to generate small masses for the active neutrinos  and its  implementation  into the standard model requires the addition of the triplet of scalars\cite{Cheng:1980qt,Gelmini:1980re,Ma:1998dx},
\begin{equation}
    \Delta=\left(\begin{array}{cc}
 \Delta^{0} & \,\frac{\Delta^{-}}{\sqrt{2}}  \\
\newline \\
\,\frac{\Delta^{-}}{\sqrt{2}} & \Delta^{--}    \end{array}\right),
\label{triplet}
\end{equation}
to  the standard Higgs doublet $H=(H^+\,,\,H^0)^T$. The  mechanism also requires explicit violation of the  lepton number occurring  in the potential by means of the non-hermitian term $\mu H^T \Delta H +H.c.$ with $\mu$ characterizing the energy scale in which lepton number is violated\cite{Ma:1998dx}. The set of constraint equations that guarantee $\Delta^0$ and $H^0$ develop vacuum expectation value (vev) different  from zero provides the relation 
\begin{equation}
    v_\Delta =\frac{\mu v^2}{\mu^2_\Delta},
    \label{SSSM}
\end{equation}
where $v_\Delta=\langle \Delta^0 \rangle$, $v=\langle H^0 \rangle$ and $\mu_\Delta$ is the mass of $\Delta$. 

The triplet $\Delta$ composes the following Yukawa term,
\begin{equation}
    {\cal L}^Y=y\bar L^C \Delta L + H.c. ,
    \label{Y-SM}
\end{equation}
where $L=(\nu_L\,,\, e_L)^T$.  When $\Delta^0$ develops vev, such term provides the following mass expression for the neutrinos 
 \begin{equation}
     m_\nu=yv_\Delta.
     \label{numass}
 \end{equation}
 
Looking at the previous expression, for natural values of $y$, it is obvious that small $v_\Delta$ leads to tiny $m_\nu$. And according to Eq. (\ref{SSSM}), a small $v_\Delta$ depends on the relation $\frac{\mu}{\mu_\Delta^2}$. Such relation leads to two interesting scenarios for $v_\Delta$ around eV. In the first one, the mechanism is performed at high energy scales with  $\mu$ and $\mu_\Delta$ belonging to the GUT scale\cite{Ma:1998dx}. Such case have theoretical motivations, since it connects tiny neutrino masses to GUT. However, it is not phenomenologically attractive, just because the triplet get very massive and can not be probed at any current and future colliders. In the second case, the mechanism is performed at low energy scales with $\mu$ belonging to the sub-keV energy range and $\mu_\Delta$ belonging to the TeV scale\cite{Li:1985hy,Lusignoli:1990yk,deSPires:2005yok}. Such scenario is more phenomenological attractive than the previous one, because  $v_\Delta$ around eV is associated to a spectrum of new scalars with  masses around TeV scales, which can be probed at the LHC. This case is very well develop  theoretically as well as phenomenologically\footnote{For some works developing the theoretical and the phenomenological aspects of the triplet model, see Refs. \cite{Freitas:2014fda,Melfo:2011nx,Arhrib:2011uy}}.

A challenge associated with this mechanism arises when we try to perform it inside gauge extensions of the standard model, as the 3-3-1 models\cite{Singer:1980sw,Pisano:1992bxx,Frampton:1992wt,Foot:1994ym}. In such models, the implementation of it is associated to a very complex scalar sector involving at least three triplets and one sextet of scalars\footnote{The first papers using the sextet to generate  masses for the neutrino into the 331RHNs are \cite{Ky:2005yq,Dong:2008sw} while the paper implementing the mechanism into the 331RHNs is \cite{Cogollo:2009yi}}. To know the profile of the spectrum of scalars associated to  such complex scalar sector is a hard, but essential, task that precedes any phenomenological approach.

In this work, we consider  the $SU(3)_C \times SU(3)_L \times U(1)_N$ model with right-handed neutrinos (331RHNs)\cite{Foot:1994ym,Montero:1992jk} and add in it one sextet of scalars to its original scalar content with the aim of implementing the type II seesaw mechanism into the model and, then, generates tiny neutrino masses. We  develop the potential  composed by such content of scalars, performing the type II seesaw mechanism in high, electroweak and low energy scales and obtain its spectrum of scalars for each case. The reason for such work lies on the fact that in order to probe the signature of the mechanism, it is essential to know the spectrum of scalars associated with it. As nice result, it is shown that when  the mechanism is performed at low energy scale, it recovers the effective THDM + triplet  at electroweak scale. For the case in which lepton number is explicitly violated at the electroweak scale, there is one specific case in which the spectrum of scalars recovers the 3HDM+triplet. For the scenario where the mechanism is performed at high energy scale, our numerical analysis show that the sextet decouples completely from the original scalar content of the model. 

This work is organized as follow: In Sec. II we present the particle content of the model where we introduce and discuss some general aspect of the sextet of scalars. In Sec. III we form the potential and present the mass matrices in the symmetric basis for all  scalars of the model. In Sec. IV we obtain the spectrum of scalars associated to lepton number  violated at low energy scale. in Sec. V we obtain numerically the spectrum of scalars for lepton number violated at the electroweak and high energy scale. In Sec. VI we present our conclusions.

\section{Particle content and representation}

In the 331RHNs model, leptons are arranged in  triplets and singlets  transforming in the following way by $SU(3)_L \times U(1)_N$ symmetry\cite{Singer:1980sw,Montero:1992jk,Foot:1994ym},
\begin{equation}
f_{aL}= \begin{pmatrix}
\nu_{a_L}     \\
\ell_{a_L}       \\
\nu^{c}_{a_R} \\
\end{pmatrix} \sim (1,3,-1/3), \quad e_{aR}\sim (1,1,-1),
\label{lp-rep}
\end{equation}
with $a=1,2,3$ representing the three SM generations of leptons. Note that the third component of the triplets carries the right-handed neutrinos.

In the Hadronic sector, anomaly cancellation requires that  one family transforms differently from the other two\cite{Frampton:1992wt}. This fact allows three possibility of arrangments for the quark families\cite{Oliveira:2022vjo}. Here we chose one where  the third generation comes in triplet and the other two come in  anti-triplet representation of $SU(3)_L$, 

\begin{eqnarray}
&&Q_{i_L} = \left (
\begin{array}{c}
d_{i} \\
-u_{i} \\
d^{\prime}_{i}
\end{array}
\right )_L\sim(3\,,\,\bar{3}\,,\,0)\,,u_{iR}\,\sim(3,1,2/3),\,\,\,\nonumber \\
&&\,\,d_{iR}\,\sim(3,1,-1/3)\,,\,\,\,\, d^{\prime}_{iR}\,\sim(3,1,-1/3),\nonumber \\
&&Q_{3L} = \left (
\begin{array}{c}
u_{3} \\
d_{3} \\
u^{\prime}_{3}
\end{array}
\right )_L\sim(3\,,\,3\,,\,1/3),u_{3R}\,\sim(3,1,2/3),\nonumber \\
&&\,\,d_{3R}\,\sim(3,1,-1/3)\,,\,u^{\prime}_{3R}\,\sim(3,1,2/3),
\label{quarks-rep} 
\end{eqnarray}
where  $i=1,2$. The primed quarks are  heavy quarks with the usual $(+\frac{2}{3}, -\frac{1}{3})$ electric charges. 

The gauge sector is composed by nine gauge bosons where four of them are the standard ones $A\,\,\,,\,\,\, W^{\pm}\,\,\,,\,\,\, Z^0$ and the other five are the typical 3-3-1 gauge bosons  $U^0 \,\,,\,\,U^{\dagger 0}\,\,,\,\,W^{\prime \pm}\,\,,\,\,Z^{\prime}$\cite{Long:1995ctv}.  For the charged and neutral currents formed with these gauge bosons and the fermions of the models, see Ref. \cite{Long:1995ctv}.

The original scalar sector of the 331RHNs involves three triplets of scalars, namely\cite{Foot:1994ym,Pal:1994ba}
\begin{eqnarray}
&&\eta = \left (
\begin{array}{c}
\eta^0 \\
\eta^- \\
\eta^{\prime 0}
\end{array}
\right ),\,\rho = \left (
\begin{array}{c}
\rho^+ \\
\rho^0 \\
\rho^{\prime +}
\end{array}
\right ),\,
\chi = \left (
\begin{array}{c}
\chi^0 \\
\chi^{-} \\
\chi^{\prime 0}
\end{array}
\right ),
\label{scalar-cont} 
\end{eqnarray}
with $\eta$ and $\chi$ transforming as $(1\,,\,3\,,\,-1/3)$, $\rho$ as $(1\,,\,3\,,\,2/3)$. This scalar content leads to the correct sequence of symmetry breaking and generate masses for all fermions except neutrinos.

With this scalar content active neutrinos may gain masses by means of effective dimension five operators\cite{Dias:2005yh}
\begin{eqnarray}
{\mathcal L}_{\nu_L}&&=	\frac{f_{l l^{\prime}}}{\Lambda}\left( \overline{L^C_l} \eta^* \right)\left( \eta^{\dagger}L_{l^{\prime}} \right)+\mbox{H.c}.
	\label{5dL}
\end{eqnarray}	
According to this operator, when $\eta^0$ develops  vev, $v_\eta$, the left-handed neutrinos develop  Majorana mass terms,
\begin{eqnarray}
(m_{\nu_L})_{l l^{\prime}}=\frac{f_{l l^{\prime}}v^2_\eta}{\Lambda}.
	\label{nuL}
\end{eqnarray}

Regarding right-handed neutrinos,  they gain masses by means of the dimension-5  operator 
\begin{eqnarray}
{\mathcal L}_{\nu_R}&&=	\frac{h_{l l^{\prime}}}{\Lambda}\left( \overline{L_{l}^C} \chi^* \right)\left( \chi^{\dagger}L_{l^{\prime}} \right)+\mbox{H.c}.
	\label{5dR}
\end{eqnarray}	
When $\chi^{\prime 0}$ develops  vev, $v_{\chi^{\prime}}$, this effective operator  provides  Majorana masses for the right-handed neutrinos,
\begin{eqnarray}
(m_{\nu_R})_{l l^{\prime}}=\frac{h_{l l^{\prime}}v^2_{\chi^{\prime}}}{\Lambda}.
	\label{nuR}
\end{eqnarray}  
Once $v_{\chi^{\prime}} > v_\eta$, then $m_{\nu_R} > m_{\nu_L}$. Thus, light right-handed neutrinos is a natural result of the model. 

These effective operators may be performed by introducing the scalar sextet\cite{Montero:2001ts} ,
\begin{equation}
    S=\frac{1}{\sqrt{2}}\left(\begin{array}{ccc}
\sqrt{2}\, \Delta^{0} & \Delta^{-} & \Phi^{0} \\
\newline \\
\Delta^{-} & \sqrt{2}\, \Delta^{--} & \Phi^{-} \\
\newline \\
\Phi^{0} & \Phi^{-} & \sqrt{2}\, \sigma^{0} \end{array}\right),
\label{sextet}
\end{equation}
with the sextet and the leptonic triplets  composing  the following Yukawa interactions
\begin{equation}
   G^{\nu}_{ab} \overline{f_{aL}}\, S^* \,(f_{bL})^c +  \mbox{H.c}.
   \label{Yukawa-sexteto}
\end{equation}
with $a,b=e,\mu,\tau$.  Observe that these interactions may lead to Majorana and Dirac mass terms involving $\nu_L$ and $\nu_R$\footnote{These cases were already implemented into the model, see Refs. \cite{Ky:2005yq,Dong:2008sw,Cogollo:2009yi}}. The potential composed with all this amount of scalars is very complex\cite{Diaz:2003dk} and its spectrum of scalars is unknown. Our idea here is to face this very hard task and obtain the spectrum of scalars  in a scenario associated to the  type II seesaw mechanism  performed at  high and low energy scales. Moreover, note that when of the $SU(3)_L \times U(1)_N$ symmetry breaking the sextet $S$ decouples into a triplet + doublet +singlet ( $S \rightarrow \Delta + \Phi +\sigma^0$ ) where $\Phi=(\Phi^0\,,\, \Phi^-)^T$. 

\section{Spectrum of scalar associated to type II seesaw mechanism: General aspects}
In this case the component  $\Phi^0$ of the sextet is assumed to be inert and, then, the Yukawa interation $G^{\nu}_{ab} \overline{f_{aL}}\, S^* \,(f_{bL})^c$  generates the following mass terms for the left-handed and right-handed neutrinos:
\begin{equation}
    m_{\nu_L}=\frac{G}{2\sqrt{2}}v_\Delta \,\,\,\,\mbox{and}\,\,\,\, m_{\nu_R}=\frac{G}{2\sqrt{2}}v_\sigma.
\label{numass}
\end{equation}
Small neutrino masses are obtained when $v_\Delta$ and $v_\sigma$ are tiny. This is obtained when lepton number is explicitly violated at low or high energy scales. Type II seesaw mechanism is a mechanism of suppression of vevs and it arises in the potential of the model.

The simplest  potential composed  with three triplet and one sextet of scalars that generates the type II seesaw mechanism is composed by the following terms,
\begin{eqnarray} 
V(\eta,\rho,\chi, S)&=&\mu_\chi^2 \chi^2 +\mu_\eta^2\eta^2
+\mu_\rho^2\rho^2+\lambda_1\chi^4 +\lambda_2\eta^4
+\lambda_3\rho^4+ \nonumber \\
&&\lambda_4(\chi^{\dagger}\chi)(\eta^{\dagger}\eta)
+\lambda_5(\chi^{\dagger}\chi)(\rho^{\dagger}\rho)+\lambda_6
(\eta^{\dagger}\eta)(\rho^{\dagger}\rho)+ \nonumber\\
&&\lambda_7(\chi^{\dagger}\eta)(\eta^{\dagger}\chi)
+\lambda_8(\chi^{\dagger}\rho)(\rho^{\dagger}\chi)+\lambda_9
(\eta^{\dagger}\rho)(\rho^{\dagger}\eta) \nonumber \\
&& \mu_S^2 Tr(S^\dagger S)+ \lambda_{10} Tr(S^\dagger S)^2 + \lambda_{11} Tr(S^\dagger SS^\dagger S) \nonumber\\
&&+ (\lambda_{12}\eta^\dagger\eta +\lambda_{13}\rho^\dagger\rho +\lambda_{14}\chi^\dagger\chi  )Tr(S^\dagger S)\nonumber\\
&&\lambda_{15} \eta^\dagger S S^\dagger \eta+ \lambda_{16} \rho^\dagger S S^\dagger \rho+\lambda_{17} \chi^\dagger S S^\dagger \chi \nonumber\\
&&- (\dfrac{M_{1}}{\sqrt{2}}\eta^T S^{\dagger} \eta  + \dfrac{M_{2}}{\sqrt{2}} \chi^T S^{\dagger} \chi  + \dfrac{f}{\sqrt{2}} \epsilon^{ijk} \eta_i \rho_j \chi_k +  \mathrm{H.c.} ).
\label{pot.typeII}
\end{eqnarray}
Of course we are aware that with three triplets and one sextet of scalars we may have a very complex potential involving many more terms than these above\cite{Tully:1998wa}. In order to have this compact potential we made use of the discrete symmetries $\eta\,,\, \rho \rightarrow -(\eta\,,\,\rho)$. 

The first thing to do here is to shift the fields around  their vevs, 
\begin{equation}
  \eta^0\,\,, \rho^0\,\,, \chi^{\prime 0}\,\,, \Delta^0\,\,, \sigma^0\,\, =\frac{1}{\sqrt{2}}\Big(v_{\eta\,, \rho\,,\chi^{\prime}\,\,, \Delta^0\,\,, \sigma^0\,\,}+R_{_{\eta\,, \rho\,,\chi^{\prime}\,\,, \Delta^0\,\,, \sigma^0\,\,}}+iI_{_{\eta\,, \rho\,,\chi^{\prime}\,\,, \Delta^0\,\,, \sigma^0\,\,, }}\Big)\,,
  \label{vevs}
\end{equation}
and then obtain the set of  equations that guarantee the potential develop a  minimum,

\begin{eqnarray}
&&\mu _{\chi }^2 + \lambda _1 v_{\chi^\prime }^2 +\frac{1}{2} \lambda _4 v_{\eta }^2 +\frac{1}{2} \lambda _5 v_{\rho }^2  +\frac{\lambda_{14}}{4}( v_{\Delta }^2 
 +v_{\sigma }^2) +\frac{\lambda _{17}}{4} v_{\sigma }^2 -\frac{1}{\sqrt{2}} \frac{f v_{\eta}v_{\rho } }{v_{\chi^{\prime}}} - M_2 v_{\sigma }=0\nonumber\\
  &&\mu _{\eta }^2+\lambda _2 v_{\eta }^2 +\frac{ \lambda _4}{2}  v_{\chi ^\prime}^2+ \frac{\lambda _6}{2} 
   v_{\rho }^2 +\frac{\lambda _{12}}{4}( v_{\Delta }^2  +v_{\sigma }^2)  +\frac{\lambda _{15}}{4} v_{\Delta }^2 -\frac{1}{\sqrt{2}} \frac{f v_{\rho } v_{\chi^\prime }}{v_\eta}- M_1 v_{\Delta }=0\nonumber\\
   && \mu _{\rho }^2 + \lambda_3 v_{\rho }^2+\frac{ \lambda _5}{2} v_{\chi^\prime }^2+\frac{ \lambda _6}{2} v_{\eta }^2 +\frac{\lambda _{13}}{4} (v_{\Delta }^2 +  v_{\sigma
   }^2)-\frac{1}{\sqrt{2}} \frac{f v_{\eta } v_{\chi^\prime }}{v_\rho}=0\nonumber\\
   && \mu _S^2+\frac{\lambda _{10}}{2}(  v_{\sigma}^2+v^2_\Delta)+\frac{\lambda _{11}}{2} v_{\Delta }^2+\frac{\lambda _{12}}{2}v_{\eta }^2+\frac{\lambda _{13}}{2}  v_{\rho }^2+\frac{\lambda _{14}}{2}  v_{\chi^\prime }^2+\frac{\lambda _{15}}{2} v_{\eta }^2- M_1 \frac{v_{\eta }^2}{v_\Delta}=0\nonumber\\
   && \mu _S^2+\frac{\lambda _{10}}{2}(  v_{\sigma}^2+v^2_\Delta)+\frac{\lambda _{11}}{2} v_{\sigma }^2+\frac{\lambda _{12}}{2}v_{\eta }^2+\frac{\lambda _{13}}{2}  v_{\rho }^2+\frac{\lambda _{14}}{2}  v_{\chi^\prime }^2+\frac{\lambda _{17}}{2} v_{\chi^{\prime} }^2- M_2 \frac{v_{\chi^{\prime} }^2}{v_\sigma}=0\nonumber
\end{eqnarray}

Observe that  the last two relations above  provide
\begin{equation}
    v_\Delta=\frac{M_1 v^2_\eta}{\mu_S^2}\,\,\,\,,\,\,\,\, v_\sigma=\frac{M_2v^2_{\chi^{\prime}}}{\mu^2_S}.
    \label{seesaw}
\end{equation}
The vevs $v_\Delta$ and $v_\sigma$ get tiny in two scenarios. In one $M_1$ and $M_2$ belong to the keV scale. This is the low scale type II seesaw mechanism.  In the other $M_1$ and $M_2$ belong to the GUT scale. This is the high scale type II seesaw. In both scenarios $v_\Delta$ lies at eV scale and $v_\sigma$ at keV scale. What is nice here is that as $\nu_L$ as well as $\nu_R$ gain masses by means of the same mechanism and both are lights. Observe that $\nu_R$ does not mix with $\nu_L$ which means that $\nu_R$ is, in fact, sterile. In summary, both neutrinos are light with their masses   obeying  a hierarchy  determined by the vevs $v_\eta$ and $v_{\chi^{\prime}}$.


Let us now obtain the mass matrices for all scalars of this model. Because $v_\Delta$ and $v_\sigma$ are tiny we neglect terms proportional to them throughout this work. We start with the  neutral complex  scalars $\Phi^0$, $\eta^{\prime}$ and $\chi^0$ and taking as basis $(\Phi,\eta^{\prime},\chi^0)$, the potential above, in conjunction with the minimum conditions, provides the following $3 \times 3$ mass matrix for these scalars,


\newcommand\scalemath[2]{\scalebox{#1}{\mbox{\ensuremath{\displaystyle #2}}}}

\begin{equation}
M_{\Phi,\chi, \eta^{\prime}}^2 = 
\left(
\scalemath{0.7}{
\begin{array}{ccc}
 \frac{M_1}{2v_\Delta} v_{\eta }^2+ -\frac{\lambda _{15}}{8} v_{\eta }^2+\frac{\lambda _{17}}{8} v_{{\chi^\prime} }^2 & -\frac{1}{2} v_{\eta }M_1& -\frac{1}{2} v_{{\chi^\prime} }M_2 \\
  -\frac{1}{2} v_{\eta }M_1 &  \frac{\sqrt{2}}{4} f \frac{v_{\rho }}{v_\eta} v_{{\chi^\prime} }+ \frac{1}{2}\left(v_{\Delta }+v_{\sigma }\right)M_1+ \frac{\lambda_7}{4} v_{\chi{^\prime}}^2  & \frac{1}{2} \left(-\frac{f v_{\rho }}{\sqrt{2}}-\frac{1}{2} \lambda _7 v_{\eta } v_{{\chi^\prime} }\right) \\
 -\frac{1}{2} v_{{\chi^\prime} }M_2 & \frac{1}{2} \left(-\frac{f v_{\rho }}{\sqrt{2}}-\frac{1}{2} \lambda _7 v_{\eta } v_{{\chi^\prime} }\right) & \frac{\sqrt{2}}{4} f \frac{v_{\rho }}{v_{{\chi^\prime} }}v_\eta + \frac{1}{2}\left(v_{\Delta }+v_{\sigma }\right)M_2+ \frac{\lambda_7}{4} v_{\eta}^2 \\
\end{array}
}
\right).
\label{mat-complex}
\end{equation}
In diagonalizing this mass matrix we must obtain one massless neutral complex  scalar that will be the Goldstone eaten by $U^0, U^{\dagger 0}$. The other two neutral complex scalars both develop mass that will be discussed below.

Let us now consider the  CP-odd fields. We organize them in the basis $(I_{\sigma},I_\Delta,I_{  \chi^\prime},I_\eta,I_\rho)$  which engenders the following mass matrix
\begin{equation}
M_I^2 = 
\left(
\begin{array}{ccccc}
 a & 0 & -\frac{M_2 v_{\chi^\prime}}{2} & 0 & 0 \\
 0 & \frac{M_1 v_\eta^2}{4 v_\Delta} & 0 & -\frac{M_1 v_\eta}{2} & 0 \\
 -\frac{M_2 v_{\chi^\prime}}{2} & 0 & \frac{f v_\eta v_\rho}{2 \sqrt{2} v_{\chi^\prime}}+M_2 v_\sigma & \frac{f v_\rho}{2 \sqrt{2}} & \frac{f v_\eta}{2 \sqrt{2}} \\
 0 & -\frac{M_1 v_\eta}{2} & \frac{f v_\rho}{2 \sqrt{2}} & \frac{f v_{\chi^\prime} v_\rho}{2 \sqrt{2} v_\eta}+M_1 v_\Delta & \frac{f v_{\chi^\prime}}{2 \sqrt{2}} \\
 0 & 0 & \frac{f v_\eta}{2 \sqrt{2}} & \frac{f v_{\chi^\prime}}{2 \sqrt{2}} & \frac{f v_{\chi^\prime} v_\eta}{2 \sqrt{2} v_\rho} \\
\end{array}
\right),
\label{mass-CP-odd}
\end{equation}
where $a=\frac{\lambda_{11} v_\Delta \left(v_\sigma^2-v_\Delta^2\right)+v_\eta^2 (2 M_1-\lambda_{15} v_\Delta)+\lambda_{17} v_{\chi^\prime}^2 v_\Delta}{8 v_\Delta}$. After diagonalizing it we must obtain two massless Goldstone bosons that will be eaten by $Z^0$ and $Z^{\prime}$. The other three massive  CP-odd scalars gain masses.

Let us now see the CP-even sector. We consider the basis $(R_{\sigma},R_\Delta,R_{  \chi^\prime},R_\eta,R_\rho)$. In this basis the mass matrix take the form
\begin{equation}
M_R^2 = \left(
\scalemath{0.6}{
\begin{array}{ccccc}
  (\frac{M_1}{4v_\Delta} -\lambda _{15}) v_{\eta }^2+\lambda _{17}  v_{{\chi^\prime} }^2& 0 & \frac{-1}{2} v_{{\chi^\prime} }M_2 & 0 & 0 \\
 0 & \frac{M_1}{4v_\Delta} v_{\eta }^2 & 0 & \frac{-1}{2} v_{\eta }M_1 & 0 \\
 \frac{-1}{2} v_{{\chi^\prime} }M_2 & 0& \frac{f v_{\eta } v_{\rho }}{2 \sqrt{2} v_{{\chi^\prime} }}+\lambda _1 v_{{\chi^\prime} }^2 & \frac{1}{2} \left(\lambda
   _4 v_{\eta } v_{{\chi^\prime} }-\frac{f v_{\rho }}{\sqrt{2}}\right) & \frac{1}{2} \left(\lambda
   _5 v_{\rho } v_{{\chi^\prime} }-\frac{f v_{\eta }}{\sqrt{2}}\right) \\
 0 & \frac{-1}{2} v_{\eta }M_1 & \frac{1}{2} \left(\lambda _4 v_{\eta } v_{{\chi^\prime} }-\frac{f v_{\rho }}{\sqrt{2}}\right) & \frac{f v_{\rho } v_{{\chi^\prime} }}{2 \sqrt{2} v_{\eta }}+\lambda
   _2 v_{\eta }^2 & \frac{1}{2} \left(\lambda _6 v_{\eta } v_{\rho }-\frac{f v_{{\chi^\prime} }}{\sqrt{2}}\right) \\
 0 & 0 & \frac{1}{2} \left(\lambda
   _5 v_{\rho } v_{{\chi^\prime} }-\frac{f v_{\eta }}{\sqrt{2}}\right) & \frac{1}{2} \left(\lambda _6 v_{\eta } v_{\rho }-\frac{f v_{{\chi^\prime} }}{\sqrt{2}}\right) & \frac{f v_{\eta } v_{{\chi^\prime}
   }}{2 \sqrt{2} v_{\rho }}+\lambda_3 v_{\rho }^2  \\
\end{array}
}
\right).
\label{mass-CP-par}
\end{equation}
On diagonalizing this mass matrix it is obtained five massive CP-even scalars and one of them must be the Higgs, $h_1$, with mass of $125$ GeV. The profile of these scalars will be discussed below.

Concerning the  charged scalars, there are six of them. The three $\chi^+, \rho^{\prime +},\Phi^+$ decouple from  the other three because of the tiny vevs.   Considering the basis ($\chi^+, \rho^{\prime +},\Phi^+$)  we have the mass matrix
\begin{equation}
M^{{+}^2}_1=
\left(
\scalemath{0.7}{
\begin{array}{ccc}
 \frac{f v_{\eta } v_{\rho }}{\sqrt{2} v_{\chi }}+M_2 v_{\sigma }+\frac{1}{2} \lambda _8 v_{\rho }^2-\frac{1}{4} \lambda _{17} v_{\sigma }^2 & \frac{1}{2} \left(\sqrt{2} f v_{\eta }+\lambda _8 v_{\rho } v_{\chi }\right) & \frac{1}{4} \lambda _{17}
   v_{\sigma } v_{\chi }-M_2 v_{\chi } \\
 \frac{1}{2} \left(\sqrt{2} f v_{\eta }+\lambda _8 v_{\rho } v_{\chi }\right) & \frac{1}{4} \left(\frac{2 \sqrt{2} f v_{\eta } v_{\chi }}{v_{\rho }}+\lambda _{16} v_{\sigma }^2+2 \lambda _8 v_{\chi }^2\right) & \frac{1}{4} \lambda _{16} v_{\rho }
   v_{\sigma } \\
 \frac{1}{4} \lambda _{17} v_{\sigma } v_{\chi }-M_2 v_{\chi } & \frac{1}{4} \lambda _{16} v_{\rho } v_{\sigma } &\frac{M_2 v_{\chi }^2}{v_{\sigma }}+\frac{1}{4} \lambda _{16} v_{\rho }^2-\frac{1}{4} \lambda _{17} v_{\chi }^2 \\
\end{array}
}
\right).
\label{mass-ch-1}
\end{equation}
On diagonalizing this matrix we must have a Goldstone dominantly given by $\chi^+$ that will be eaten by the gauge boson $W^{\prime +}$.

The other three charge scalars, arranged in the basis $(\eta^+\,,\, \rho^+\,,\,\Delta^+)$  have the following mass  matrix 
\begin{equation}
M_2^{{+}^2}=\left(
\scalemath{0.8}{
\begin{array}{ccc}
 \frac{v_{\rho } \left(\sqrt{2} f v_{\chi }+\lambda _9 v_{\eta } v_{\rho }\right)}{2 v_{\eta }}+M_1 v_{\Delta }-\frac{1}{4} \lambda _{15} v_{\Delta }^2 & \frac{1}{2} \left(\sqrt{2} f v_{\chi }+\lambda _9 v_{\eta } v_{\rho }\right) & \frac{1}{4}
   \lambda _{15} v_{\Delta } v_{\eta }-M_1 v_{\eta } \\
 \frac{1}{2} \left(\sqrt{2} f v_{\chi }+\lambda _9 v_{\eta } v_{\rho }\right) & \frac{1}{4} \left(\frac{2 v_{\eta } \left(\sqrt{2} f v_{\chi }+\lambda _9 v_{\eta } v_{\rho }\right)}{v_{\rho }}+\lambda _{16} v_{\Delta }^2\right) & \frac{1}{4}
   \lambda _{16} v_{\Delta } v_{\rho } \\
 \frac{1}{4} \lambda _{15} v_{\Delta } v_{\eta }-M_1 v_{\eta } & \frac{1}{4} \lambda _{16} v_{\Delta } v_{\rho } & \frac{M_1 v_{\eta }^2}{v_{\Delta }}-\frac{1}{4} \lambda _{15} v_{\eta }^2+\frac{1}{4} \lambda _{16} v_{\rho }^2  \\
\end{array}
}
\right).
\label{mass-ch-2}
\end{equation}

On diagonalizing this mass matrix we must have a Goldstone that is dominantly a mixture of $\eta^+$ and $\rho^+$ and will be eaten by $W^{\pm}$.

There is only one  doubly charged scalar $\Delta^{++}=H^{++}$ and it gains the following mass expression $m^2_{H^{\pm\pm}}=\frac{M_1}{2v_\Delta}v^2_\eta -\frac{\lambda_{15}}{4}v^2_\eta+\frac{\lambda_{16}}{4}v^2_\rho -\frac{\lambda_{11}}{4}v^2_\Delta$.

Those are the textures of the mass matrices associated to the type II seesaw mechanism in the 331RHNs.  In what follow we diagonalize these mass matrices with the aim of obtaining  the spectrum of scalars. We do this for the cases of $M_1$ and $M_2$ belonging to low, electroweak and high energy scales
.

Observe that there is a third energy scale parameterized by $f$. This parameter is related to the explicitly breaking of a global $U(1)_X$ symmetry that was recognized in \cite{Pal:1994ba} to be the Peccei-Quinn symmetry.  A deep investigation of the potential with three triplet done in  Ref. \cite{Pinheiro:2022bcs} suggested that the range of values for the parameter $f$ that is phenomenologicaly interesting lies at electroweak scale or lower. We, then, restrict our analysis for the case $f$ is lower than the electroweak scale $(f< v_\eta\,,\, v_\rho)$ and for the case $(f=v_\eta\,,\, v_\rho)$.

\section{Spectrum of scalars associated to the low scale type II seesaw mechanism and $f< v_\eta\,,\, v_\rho$}
Low scale type II seesaw mechanism is characterized by lepton number being explicitly violated at low energy scale more precisely around sub-keV scale which means to take  $M_1=M_2=M\sim 10^{-7}$ GeV. This scale is interesting because $v_\Delta $ around eV  and $v_\sigma$ around keV get associated to physics at TeV scale. In practical terms this scenario is characterized by  $M_1\,,\,M_2 \ll v_{\chi^{\prime}}\,,\,v_\eta\,,\, v_\rho \,,\, f$. We also take $v_{\chi^{\prime}}>v_\eta\,,\, v_\rho$ and $f< v_\eta\,,\, v_\rho$.

Let us first focus on the mass matrix   $M_{\Phi,\chi, \eta^{\prime}}^2$ given in Eq. (\ref{mat-complex}). According to the approximations, it takes the form,

\[
M_{\Phi,\chi, \eta^{\prime}}^2 \approx 
\left(
\scalemath{0.7}{
\begin{array}{ccc}
 \frac{M}{2v_\Delta} v_{\eta }^2+ -\frac{\lambda _{15}}{8} v_{\eta }^2+\frac{\lambda _{17}}{8} v_{{\chi^\prime} }^2 & 0& 0 \\
 0 &  \frac{\sqrt{2}}{4} f \frac{v_{\rho }}{v_\eta} v_{{\chi^\prime} }+ \frac{\lambda_7}{4} v_{\chi{^\prime}}^2  & \frac{1}{2} \left(-\frac{f v_{\rho }}{\sqrt{2}}-\frac{1}{2} \lambda _7 v_{\eta } v_{{\chi^\prime} }\right) \\
 0 & \frac{1}{2} \left(-\frac{f v_{\rho }}{\sqrt{2}}-\frac{1}{2} \lambda _7 v_{\eta } v_{{\chi^\prime} }\right) & \frac{\sqrt{2}}{4} f \frac{v_{\rho }}{v_{{\chi^\prime} }}v_\eta + \frac{\lambda_7}{4} v_{\eta}^2 \\
\end{array}
}
\right).
\]
Observe that  $H_2 \approx \Phi^0$ decoupled from  $\eta^{\prime}$ and $\chi$ and gains    mass  $m^2_{H_2} \approx \frac{M}{2v_\Delta} v_{\eta }^2 -\frac{\lambda _{15}}{8} v_{\eta }^2+\frac{\lambda _{17}}{8} v_{{\chi^\prime} }^2$  at the  3-3-1 scale which can be probed at LHC.

Diagonalizing the remaining $2 \times 2$ block matrix we have a decoupling where   $\chi$  is the Goldstone eaten by the non-hermitian gauge bosons  $U^{0 \dagger}$ and $U^0$, while $H_1\approx {\eta^{\prime }}$ has mass  given by $m^2_{H_1} \approx \frac{\lambda_7}{4}(v_{\chi^{\prime}}^2+ v^2_\eta)+\frac{fv_\rho}{4}(\frac{v_{\chi^{\prime}}}{v_\eta}+\frac{v_\eta}{v_{\chi^{\prime}}})$   which lies at the 3-3-1 scale.

According to the assumptions the mass matrix for the  CP-odd scalars given in Eq. (\ref{mass-CP-odd}) get the expression,
\begin{equation}
M_I^2 \approx 
\left(
\begin{array}{ccccc}
 \lambda_{17} v_{\chi^\prime}^2 +v_\eta^2 (\frac{ M}{4v_\Delta}-\frac{\lambda_{15}}{8} ) & 0 & 0 & 0 & 0 \\
 0 & \frac{M v_\eta^2}{4 v_\Delta} & 0 & 0 & 0 \\
 0 & 0 & \frac{f v_\eta v_\rho}{2 \sqrt{2} v_{\chi^\prime}} & \frac{f v_\rho}{2 \sqrt{2}} & \frac{f v_\eta}{2 \sqrt{2}} \\
 0 & 0 & \frac{f v_\rho}{2 \sqrt{2}} & \frac{f v_{\chi^\prime} v_\rho}{2 \sqrt{2} v_\eta} & \frac{f v_{\chi^\prime}}{2 \sqrt{2}} \\
 0 & 0 & \frac{f v_\eta}{2 \sqrt{2}} & \frac{f v_{\chi^\prime}}{2 \sqrt{2}} & \frac{f v_{\chi^\prime} v_\eta}{2 \sqrt{2} v_\rho} \\
\end{array}
\right).
\end{equation}

Here also the particles belonging to the sextet, $A_3 \approx I_\sigma$ and $A_2 \approx I_\Delta$,  decouple from the particles belonging to the triplets,  $I_{\chi^{\prime}}$, $I_\eta$ and $I_\rho$, in such a way that after diagonalization we get two null eigenvalues, which correspond to the eigenstates $A=I_{\chi^{\prime}}$ and $A^{\prime}= \frac{v_\eta}{\sqrt{v^2_\eta + v^2_\rho}} I_\eta -\frac{v_\rho}{\sqrt{v^2_\eta + v^2_\rho}}I_\rho$, and  a massive one with mass expression given by  $m^2 _{A_1}=\frac{f v_{\chi^{\prime}}}{4}(\frac{v_\eta v_\rho}{v^2_{\chi^{\prime}}}+\frac{v_\eta}{v_\rho}+\frac{v_\rho}{v_\eta})$ corresponding to the eigenstate  $A_1= \frac{v_\rho}{\sqrt{v^2_\eta + v^2_\rho}} I_\eta +\frac{v_\eta}{\sqrt{v^2_\eta + v^2_\rho}}I_\rho$.  See that   the parameter $f$ determine  $m_{A_1}$ which means that, in this particular case,  $A_1$ get mass at electroweak scale. This is a very nice result since any bound  on $m_{A_1}$ implies in a  constraint on $f$. This point was already  addressed in Ref. \cite{Pinheiro:2022bcs} where it was found that the decay of the standard Higgs in couple of $A_1$ implies $f> 0.8$ GeV. 

Concerning the other two CP-odd scalars, their  masses are given by $m^2_{A_3} \approx \lambda_{17} v_{\chi^\prime}^2 +v_\eta^2 (\frac{ M}{4v_\Delta}-\frac{\lambda_{15}}{8} )$ and $ m^2_{A_2}\approx \frac{M v_\eta^2}{4 v_\Delta}$.   The first has mass at the 3-3-1 scale while the second, because   $\frac{M}{v_\Delta} \approx 10^2$,  has mass at the electroweak scale. In summary, we have two Goldstones and three massive  CP-odd scalars with two of them having masses at  the electroweak scale and only one with mass at  the 3-3-1 scale.

For the other five CP-even scalars, their mass matrix in Eq. (\ref{mass-CP-par}) becomes,
\[
M_R^2 \approx \left(
\scalemath{0.6}{
\begin{array}{ccccc}
  (\frac{M}{4v_\Delta} -\lambda _{15}) v_{\eta }^2+\lambda _{17}  v_{{\chi^\prime} }^2& 0 & 0 & 0 & 0 \\
 0 & \frac{M}{4v_\Delta} v_{\eta }^2 & 0 & 0 & 0 \\
 0 & 0& \frac{f v_{\eta } v_{\rho }}{2 \sqrt{2} v_{{\chi^\prime} }}+\lambda _1 v_{{\chi^\prime} }^2 & \frac{1}{2} \left(\lambda
   _4 v_{\eta } v_{{\chi^\prime} }-\frac{f v_{\rho }}{\sqrt{2}}\right) & \frac{f v_{\eta }}{2 \sqrt{2}} \\
 0 & 0 & \frac{1}{2} \left(\lambda _4 v_{\eta } v_{{\chi^\prime} }-\frac{f v_{\rho }}{\sqrt{2}}\right) & \frac{f v_{\rho } v_{{\chi^\prime} }}{2 \sqrt{2} v_{\eta }}+\lambda
   _2 v_{\eta }^2 & \frac{1}{2} \left(\lambda _6 v_{\eta } v_{\rho }-\frac{f v_{{\chi^\prime} }}{\sqrt{2}}\right) \\
 0 & 0 & \frac{f v_{\eta }}{2 \sqrt{2}} & \frac{1}{2} \left(\lambda _6 v_{\eta } v_{\rho }-\frac{f v_{{\chi^\prime} }}{\sqrt{2}}\right) & \frac{f v_{\eta } v_{{\chi^\prime}
   }}{2 \sqrt{2} v_{\rho }}+\lambda_3 v_{\rho }^2 \\
\end{array}
}
\right).
\]

Note that $h_2 \approx R_\Delta$ and $h_4 \approx R_\sigma$ decouples from the other CP-even scalars and gain mass $m^2_{h_2}\approx (\frac{M}{4v_\Delta} -\lambda _{15}) v_{\eta }^2+\lambda _{17}  v_{{\chi^\prime} }^2$ and  $m^2_{h_4}\approx \frac{M}{4v_\Delta} v_{\eta }^2$. The first lie at TeV scale while the second, as discussed above, lies at the electroweak scale. The other three CP-even scalars mix among themselves and their masses are obtained from the diagonalization of $3\times 3$ sub-matrix in $M^2_R$. This mass matrix  was investigated in Ref. \cite{Pinheiro:2022bcs} and as result for the specific case $v_\eta=v_\rho=v$ we have that $h_3 \approx R_\chi$ decouples from $R_\eta$ and $R_\rho$ and has mass $m^2_{h_3} \approx \lambda_1 v^2_{\chi^{\prime}}$.  The other two CP-even, $R_\eta$ and $R_\rho$, mix to form $h_1$ and $h_5$,
\begin{eqnarray}
&&h_1\cong\frac{\lambda_4}{\sqrt{\lambda_4^2+\lambda_5^2}}R_\eta +\frac{\lambda_5}{\sqrt{\lambda_4^2+\lambda_5^2}}R_\rho ,\nonumber \\
  &&
  h_5\cong-\frac{\lambda_5}{\sqrt{\lambda_4^2+\lambda_5^2}}R_\eta +\frac{\lambda_4}{\sqrt{\lambda_4^2+\lambda_5^2}}R_\rho .
  \label{S1}
\end{eqnarray}
with,
\begin{equation}
   m^2_{h_1}=(\lambda_2+\frac{\lambda_6}{2})v^2\,\,\,,\,\,\, m^2_{h_5}=(\lambda_2-\frac{\lambda_6}{2})v^2+fv_{\chi^{\prime}}
    \label{S2}
\end{equation}
We recognize $h_1$ as the scalar that will play the role of the Higgs and $h_5$ as the second Higgs with mass around the electroweak scale for tiny $f$.

Now let us see the simply charged scalars. According to the approximation done here the mass matrix in Eq. (\ref{mass-ch-1}) get the  form
\[
M_1^{{+}^2}=
\left(
\scalemath{0.7}{
\begin{array}{ccc}
 \frac{f v_{\eta } v_{\rho }}{\sqrt{2} v_{\chi }}+\frac{1}{2} \lambda _8 v_{\rho }^2 & \frac{1}{2} \left(\sqrt{2} f v_{\eta }+\lambda _8 v_{\rho } v_{\chi }\right) & 0 \\
 \frac{1}{2} \left(\sqrt{2} f v_{\eta }+\lambda _8 v_{\rho } v_{\chi }\right) & \frac{1}{4} \left(\frac{2 \sqrt{2} f v_{\eta } v_{\chi }}{v_{\rho }}+2 \lambda _8 v_{\chi }^2\right) &  0 \\
 0& 0 &\frac{M v_{\chi }^2}{v_{\sigma }}+\frac{1}{4} \lambda _{16} v_{\rho }^2-\frac{1}{4} \lambda _{17} v_{\chi }^2 \\
\end{array}
}
\right).
\]
On diagonalizing this matrix, all the three eigenstates decouple. We have that $\chi^+$  is the Goldstone eaten by $W^{\prime +}$. The others charged scalars are $H^+_1 \approx \Phi^+$ has mass $m^2_{H_2^+}\approx \frac{M v_{\chi }^2}{v_{\sigma }}+\frac{1}{4} \lambda _{16} v_{\rho }^2-\frac{1}{4} \lambda _{17} v_{\chi }^2$ and $H^+_1 \approx \rho^{\prime +} $   has mass  $m^2_{H_1^+}\approx \frac{1}{2}(f v_{\eta}+\lambda_8 v_\rho v_{\chi^{\prime}})(\frac{v_{\chi^{\prime}}}{v_\rho} +\frac{v_\rho}{v_{\chi^{\prime}}})$. Both charged scalars have mass lying at 3-3-1 scale which may be probed at present colliders as LHC.

For the other mass  matrix in Eq. (\ref{mass-ch-2}), we have

\[M_2^{{+}^2}\approx \left(
\scalemath{0.8}{
\begin{array}{ccc}
 \frac{v_{\rho } \left(\sqrt{2} f v_{\chi }+\lambda _9 v_{\eta } v_{\rho }\right)}{2 v_{\eta }} & \frac{1}{2} \left(\sqrt{2} f v_{\chi }+\lambda _9 v_{\eta } v_{\rho }\right) & 0 \\
 \frac{1}{2} \left(\sqrt{2} f v_{\chi }+\lambda _9 v_{\eta } v_{\rho }\right) & \frac{1}{4} \left(\frac{2 v_{\eta } \left(\sqrt{2} f v_{\chi }+\lambda _9 v_{\eta } v_{\rho }\right)}{v_{\rho }}+\lambda _{16} v_{\Delta }^2\right) & 0 \\
 0 & 0 & \frac{M v_{\eta }^2}{v_{\Delta }}-\frac{1}{4} \lambda _{15} v_{\eta }^2+\frac{1}{4} \lambda _{16} v_{\rho }^2 \\
\end{array}
}
\right).
\]
On diagonalizing it we get one Goldstone $G_h^+=-\frac{v_\rho}{\sqrt{v^2_\eta + v^2_\rho}}\rho^+ +\frac{v_\eta}{\sqrt{v^2_\eta + v^2_\rho}} \eta^+ +\frac{v_\Delta}{\sqrt{v^2_\eta + v^2_\rho}} \Delta^+  \approx \frac{v_\eta}{\sqrt{v^2_\eta + v^2_\rho}} \eta^+ -\frac{v_\rho}{\sqrt{v^2_\eta + v^2_\rho}}\rho^+ $, which is eaten by $W^{\pm}$, and the predominantly $h^+_2 \approx \Delta^+$ with mass $m^2_{h_2^+} \approx \frac{M v_{\eta }^2}{v_{\Delta }}-\frac{1}{4} \lambda _{15} v_{\eta }^2+\frac{1}{4} \lambda _{16} v_{\rho }^2$ which lie at electroweak scale and may be probed at the LHC. There is also  $h_1^+=\frac{v_\eta}{\sqrt{v^2_\eta + v^2_\rho}}\rho^+ +\frac{v_\rho}{\sqrt{v^2_\eta + v^2_\rho}} \eta^+ $ with mass $m_{h_1^{\pm}}^2=\frac{1}{2}(f v_{\chi^{\prime}}+\lambda_9 v_\eta v_\rho)(\frac{v_\eta}{v_\rho}+\frac{v_\rho}{v_\eta}) $. Observe that this charged scalar has mass at the electroweak scale. Concerning the doubly charged scalar presented above, observe that , here, its mass also lies at the electroweak scale, too. 

Summarizing,  we saw that  in the type II seesaw mechanism at low energy scale with tiny $f$  the set of scalars composed by  $h_1$ (the Higgs),  $h_5$ (the second Higgs),  $A_1$ (pseudoscalar) and $h_1^+$ characterizes the THDM, while the set of scalars composed by  $A_2\approx I_\Delta$, $h_4 \approx R_\Delta$, $h_2^+\approx \Delta^+$ and $H^{++}\approx \Delta^{++}$  characterizes the triplet of scalars. These two set of scalars  gain masses at the electroweak and together  recover the so-called  THDM + triplet of scalars.  The  other set of scalars gain masses at the 3-3-1 scale. The whole scenario is phenomenological attractive once all scalars gain masses among electroweak and TeV scale which can be probed  at present or future colliders. 

Concerning the phenomenology, in part it is determined by the THDM+triplet whose phenomenological aspects have been intensively studied in many papers, see \cite{Arhrib:2011uy,Melfo:2011nx} and references therein. For the search of the  triplet at the LHC, see Refs. \cite{Freitas:2014fda,Cogollo:2019mbd,Primulando:2019evb,Ashanujjaman:2021txz}. For collider search of these scalars in the context of the 331RHNs, see \cite{deSousaPires:2018fnl,Ferreira:2019qpf} . There is also  a second Higgs with mass close to the standarg Higgs whose implications deserve care. The other part of the scalar spectrum also has interesting phenomenological implications mainly in processes involving flavor changing neutral currents\cite{Long:1999ij,Cogollo:2012ek,Okada:2016whh,Huitu:2019kbm,Oliveira:2022dav}.


To finish, it was pointed out in Ref. \cite{Pinheiro:2022bcs}  that the case  $f=v_{\chi^{\prime}}$ and $f\gg  v_{\chi^{\prime}}$ are not attractive phenomenologicaly once push up the masses of all scalars belonging to the triplets $\eta$, $\rho$ and $\chi$. We do not address such cases here.

\section{Spectrum of scalars associated to the type II seesaw mechanism varying $M_1\,,\,M_2\,,\, f$ arbitrarily: a numerical approach }

 In this section, we present the results of our analysis for the masses of the scalars  with   $M_1$ and $M_2$ varying from sub-keV up to thousands of TeV and $f$ varying from low up to  the electroweak symmetry breaking (EWSB) scale which we refer as $v$.

We focus in cases that are attractive phenomenologically. For pedagogical reason, we  break our analysis into two blocks. In one block we  study the case in which $v_\sigma$ and $v_\Delta$ vary from eV  up to sub-MeV scale while in the second block they can vary from sub-MeV up to electroweak scale. We also assume two substructures related to the parameters $M_1\,\,,\,M_2$ inside each block. 

Our results are summarized in the figures bellow. Using 1 million of points for each energy regime, we considered only the solutions in which the mass of all scalars are positive definite. The blue dots represents all possible positive definite masses of the scalars at tree-level and the red dots represents the positive definite masses of the scalars with the Higgs-like particle mass ($m_{h_1}$) lying between $100$ GeV and $200$ GeV.
However, before proceeding to the analysis itself, we must discuss about the number of physical massive scalar particles of the studied model. In total there are 5 CP-even scalars, $h_{(1,\,2, \, 3,\,4,\,5)}$,
where one of them is the Higgs-like particle ($h_1$). There are 3 CP-odd scalars, $A_{(1,\,2,\,3)}$, and 2 complex neutral scalars $H_{(1,\,2)}$. There are four singly charged scalars. Among them two are associated with the 331 typical energy scale, $H_{(1,\,2)}^\pm$, and the other two are  $h_{(1,\,2)}^\pm$. The last massive particle is the doubly charged scalar. Then, there are 15 massive scalars to be analyzed.

Here, we try to understand how sensitive are the masses of these scalars to the variation of the vev's $v_\sigma$ and $v_\Delta$. For example,  $H_1^\pm$ and $H_1$ are dominantly composed by $\Phi$ components. We will see  that tiny $v_\sigma$ leaves the mass of such scalars strictly associated with the 331 scale or heavier. However, for high values of the vev's $v_\Delta$ and $v_\sigma$, and not imposing any hierarchy among $v_\Delta$ and $M_1$, $M_2$, there is a reduction in the masses of some scalars  to hundreds of GeV. In addition, there is a set of  particles that always belong to the 331 energy scales for all mentioned hierarchies. For example, the CP-even scalars $h_2$ and $h_3$  are a composition of $R_\sigma$ and $R_{\chi^\prime}$. Typically they are degenerated in masses and their mixing will depend on the scale of the $M$'s. The $h_4$ scalar is dominantly composed by $R_\Delta$. The particles  $h_1$ and $h_5$ are almost a perpendicular composition of $\rho$ and $\eta$ (such mixing depends on the scale of $f$). In the pseudoscalar sector, the CP-odd scalar $A_3$  is predominantly composed by $I_\sigma$. The complex neutral scalar $H_2$ is dominantly composed by $\Phi^0$ and the charged scalar \textbf{ $H_2^\pm$ by $\Phi^\pm$ }. The particles cited above  have masses typically at the 331 scale and they  will not be mentioned in the following analysis, unless we say something explicitly about them.

As it will be see in the end, the scalar sector of the 331+sextet, does not matter the energy regime of  $v_\Delta$ and $v_\sigma$ , it recovers effectively the phenomenology of the THDM+triplet.  However, for high values of $M_1$ and $M_2$, the triplet $\Delta$ totally decouples from the particles of the THDM. The singlet $\sigma$ and the doublet $\Phi$ inside the sextet decouples from the EWSB particles, mixing only with the typical 331 scalars, as described above.  For the other side, the components of the triplet inside the sextet ($\Delta$) mixes with some of the components of the triplets $\eta$ and $\rho$, simulating the simple case of the THDM+triplet.  The number of phenomenologically viable particles are associated directly with the values of the parameters $M_1$, $M_2$  and $f$.  For example, for high values of $f$ and low values of $M_1,\;M_2$, all the particles of the THDM decouples from the scalar sector, remaining a Higgslike particle and the components of the triplet $\Delta$, recovering effectively the well studied case of the SM+triplet scenario. And, as discussed briefly above, if $f$ remains at EWSB energy scales and $M_1,\;M_2$ goes to very high energies, all the masses of the particles of the sextet skyrocket in the same proportion of such parameters and decouple from the other particles of the model.

\subsection{$v_\sigma$ and $v_\Delta \ll v$}
For this case, we are letting $f$ to vary between $10^{-6}$ GeV and $10^{3}$ GeV. The vev's of the scalars $\eta$ and $\rho$ as  $v_\rho = \sqrt{246.11^2\;\mathrm{GeV}^2-v_\eta^2}$ and  $0$ GeV$<v_\eta<246.11$ GeV and
the vev of the triplet $\chi$ is varying between $10^3\;\mathrm{GeV}<v_{\chi^\prime}<10^{10}$ GeV. The vev of the triplet of the sextet ($v_\Delta$) is varying as $10^{-12}$ GeV$<v_\Delta <10^{-4}$ GeV and the vev of the singlet of the sextet ($v_\sigma$) is varying between 
$10^{-7}$ GeV$<v_\sigma <10^{-4}$ GeV. They always respect the hierarchy among the vev's as $v_\Delta < v_\sigma< v=246.11$ GeV$<v_{\chi^\prime}$. Another important hierarchy that we are considering in our analysis is that $v_\Delta < M_1,\;M_2$ for all cases studied in this block.
To finish, all quartic couplings are varying between $-\sqrt{4\pi}$ and $\sqrt{4\pi}$. All these parameters will vary between this interval no matter the energy regime of $M_1$ and $M_2$.

As discussed above, we are varying the parameters $M_1$ and $M_2$ between two energy regimes and consequently two different hierarchy among these parameters and $f$.
The first hierarchy among them is $M_1,M_2< f$, such that $10^{-12}\;\mathrm{GeV}<M_1,M_2<10^{-6}\;\mathrm{GeV}$. The second hierarchy is
$M_1,M_2\sim f$, such that $10^{-6}\;\mathrm{GeV}<M_1,M_2<10^{3}$ GeV.  

\subsubsection{$M_1,M_2<f\sim v$}

\begin{figure}[ht] 
  \begin{subfigure}[b]{0.5\linewidth}
    \centering\includegraphics[width=0.75\linewidth]{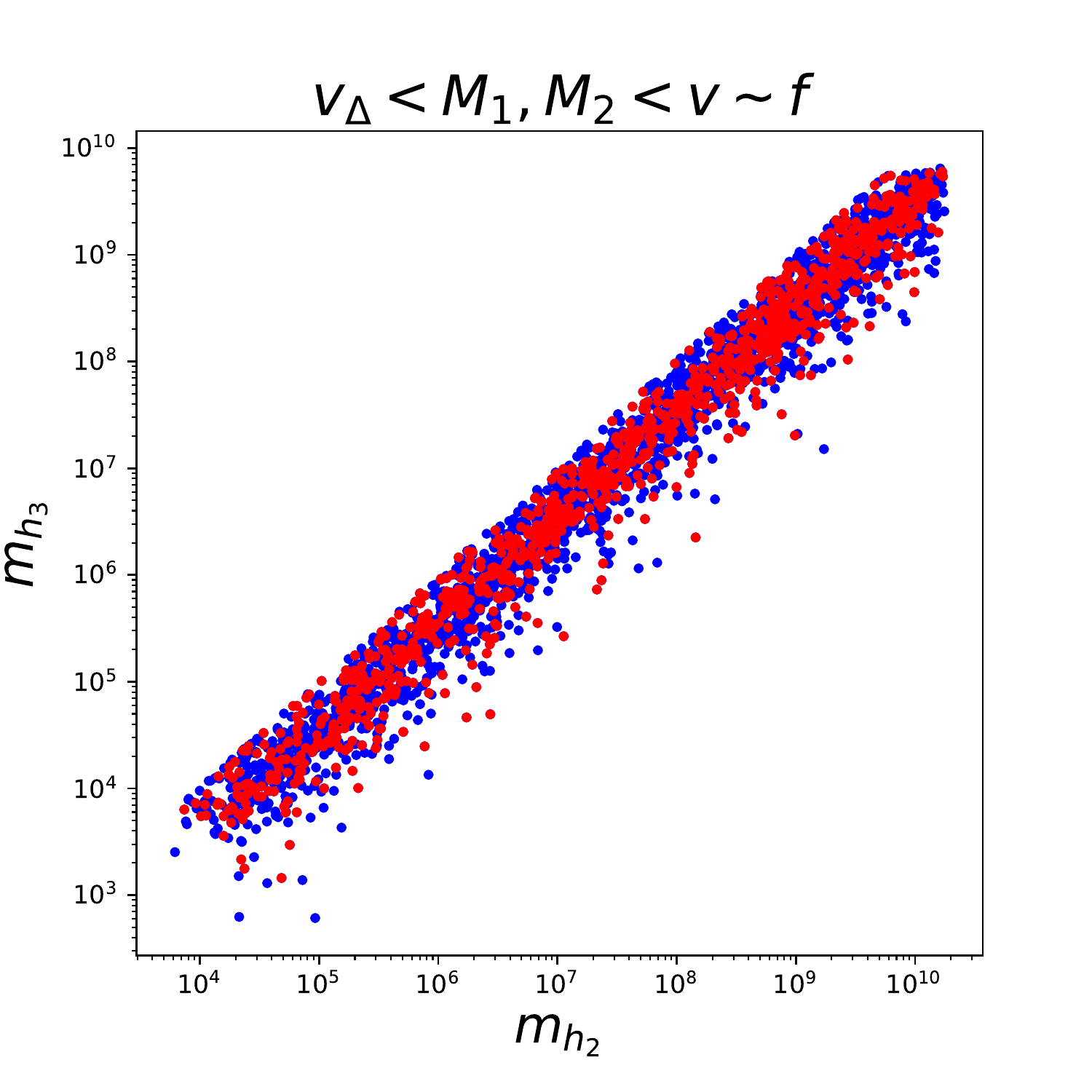} 
    \label{fig1b:a} 
    \vspace{4ex}
  \end{subfigure}
  \begin{subfigure}[b]{0.5\linewidth}
    \centering
\includegraphics[width=0.75\linewidth]{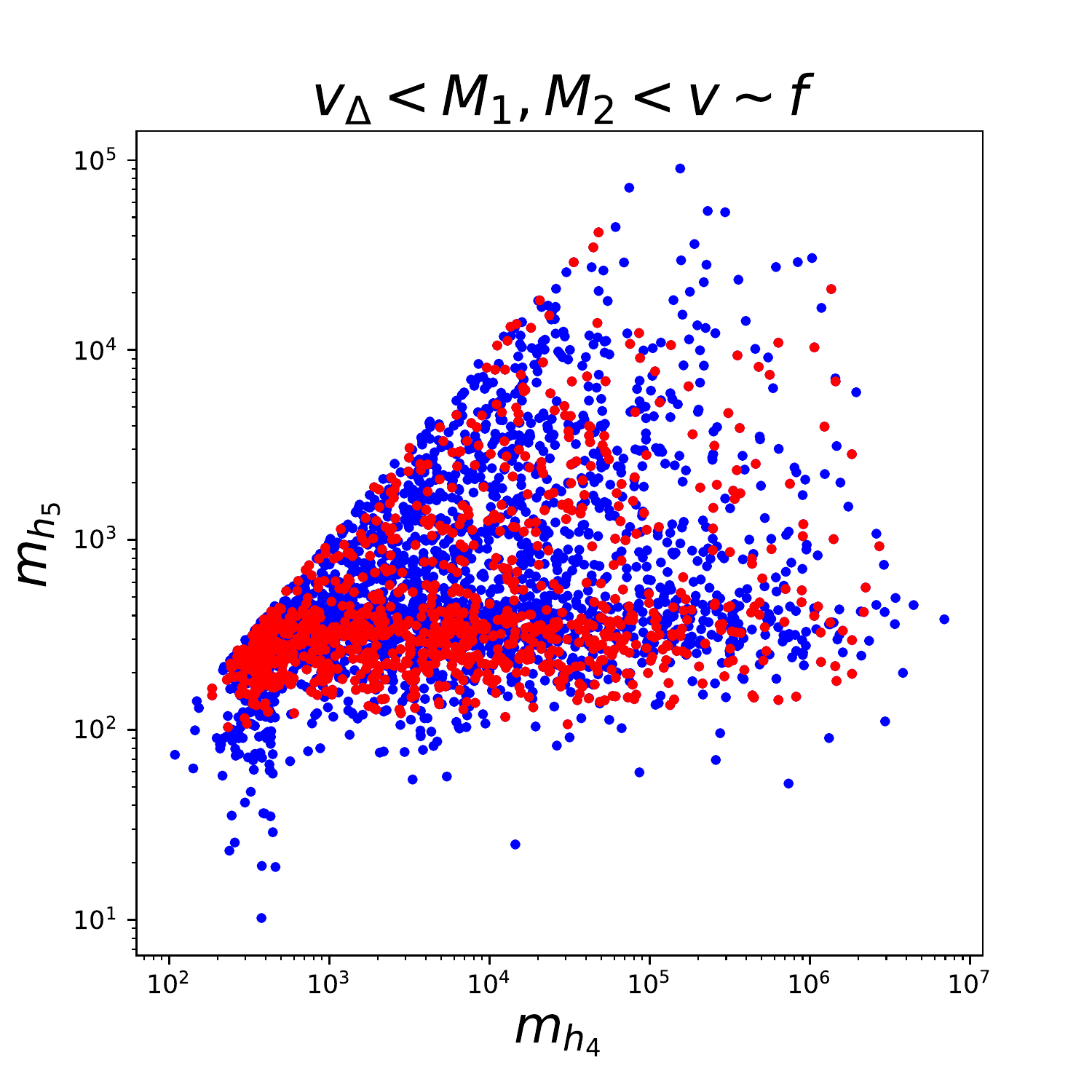} 
    \label{fig1b:b} 
    \vspace{4ex}
  \end{subfigure} 
  \begin{subfigure}[b]{0.5\linewidth}
    \centering    \includegraphics[width=0.75\linewidth]{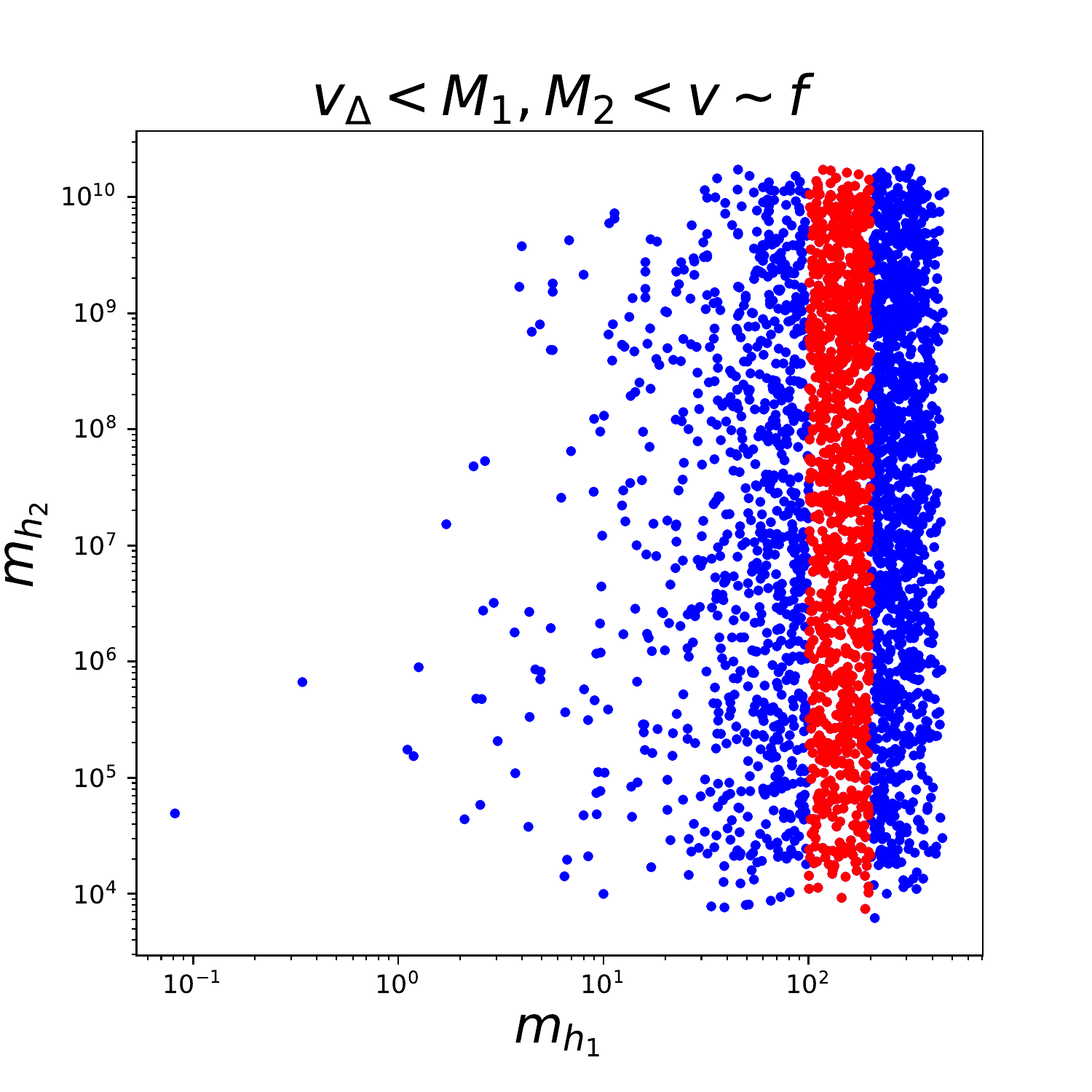}
    \label{fig1b:c} 
  \end{subfigure}
\caption{CP even at $M_1,M_2<f\sim v$. The blue dots represents all possible positive definite masses of the scalars at tree-level and the red dots represents the positive definite masses of the scalars with the Higgs-like particle mass ($m_{h_1}$) lying between $100$ GeV and $200$ GeV(See the text).}
  \label{fig1b} 
\end{figure}

 In the CP-even sector, as can be seem in FIG. \ref{fig1b}, among the five particles mentioned above, two of them have masses typically of the EWSB energy scale ($h_1$ and $h_5$) and they  are a perpendicular composition of $R_\eta$ and $R_\rho$. The correct mass for the Higgs ($h_1$) imposes that the scalar $h_5$ must be slightly heavier than $h_1$. The same for the interesting scalar  $h_4$,  that is almost predominantly composed by $R_\Delta$, and its mass can vary between $10^2$ GeV and $10^7$ GeV.

\begin{figure}[ht] 
  \begin{subfigure}[b]{0.5\linewidth}
    \centering    \includegraphics[width=0.75\linewidth]{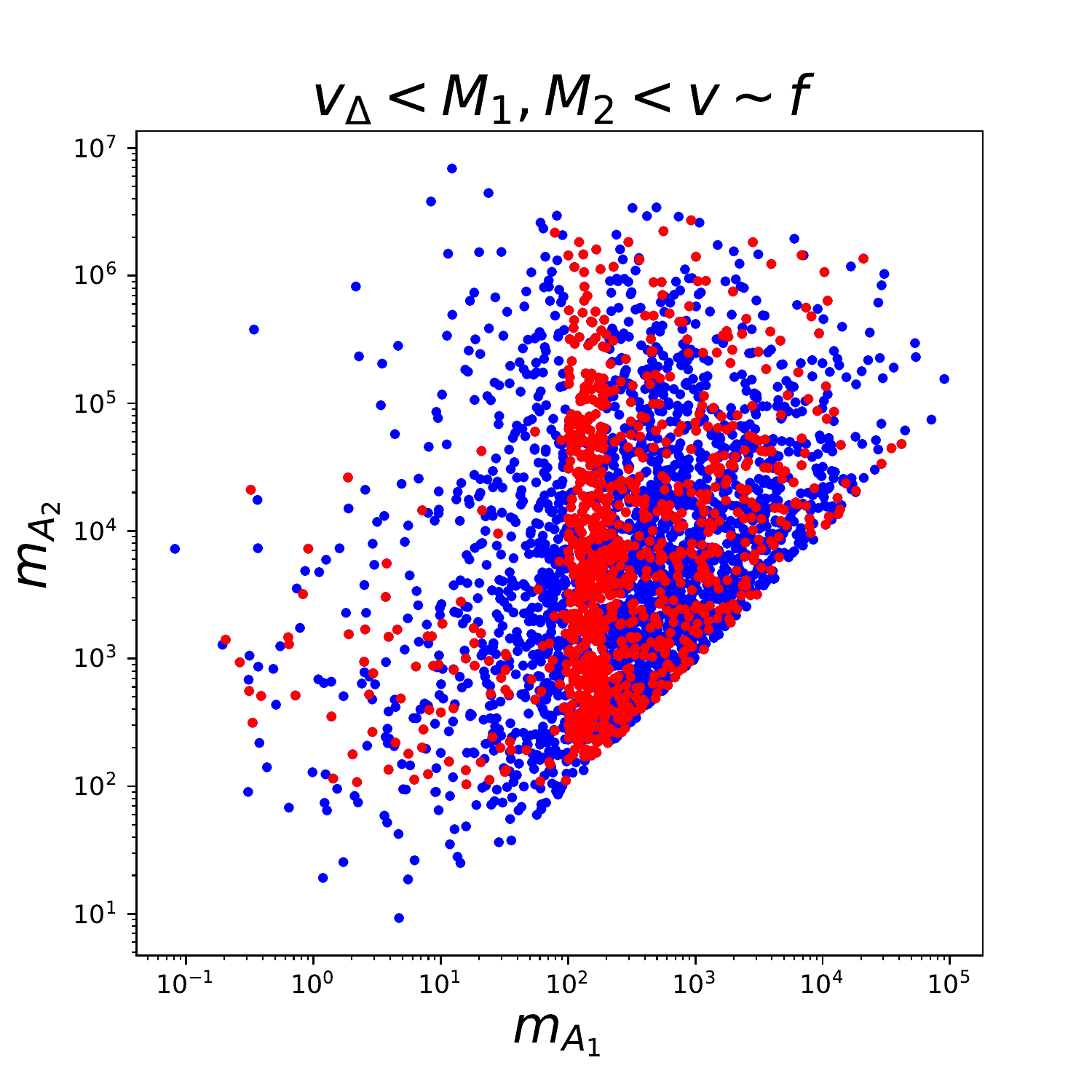}    
    \label{fig4b:a} 
    \vspace{4ex}
  \end{subfigure}
  \begin{subfigure}[b]{0.5\linewidth}
    \centering
\includegraphics[width=0.75\linewidth]{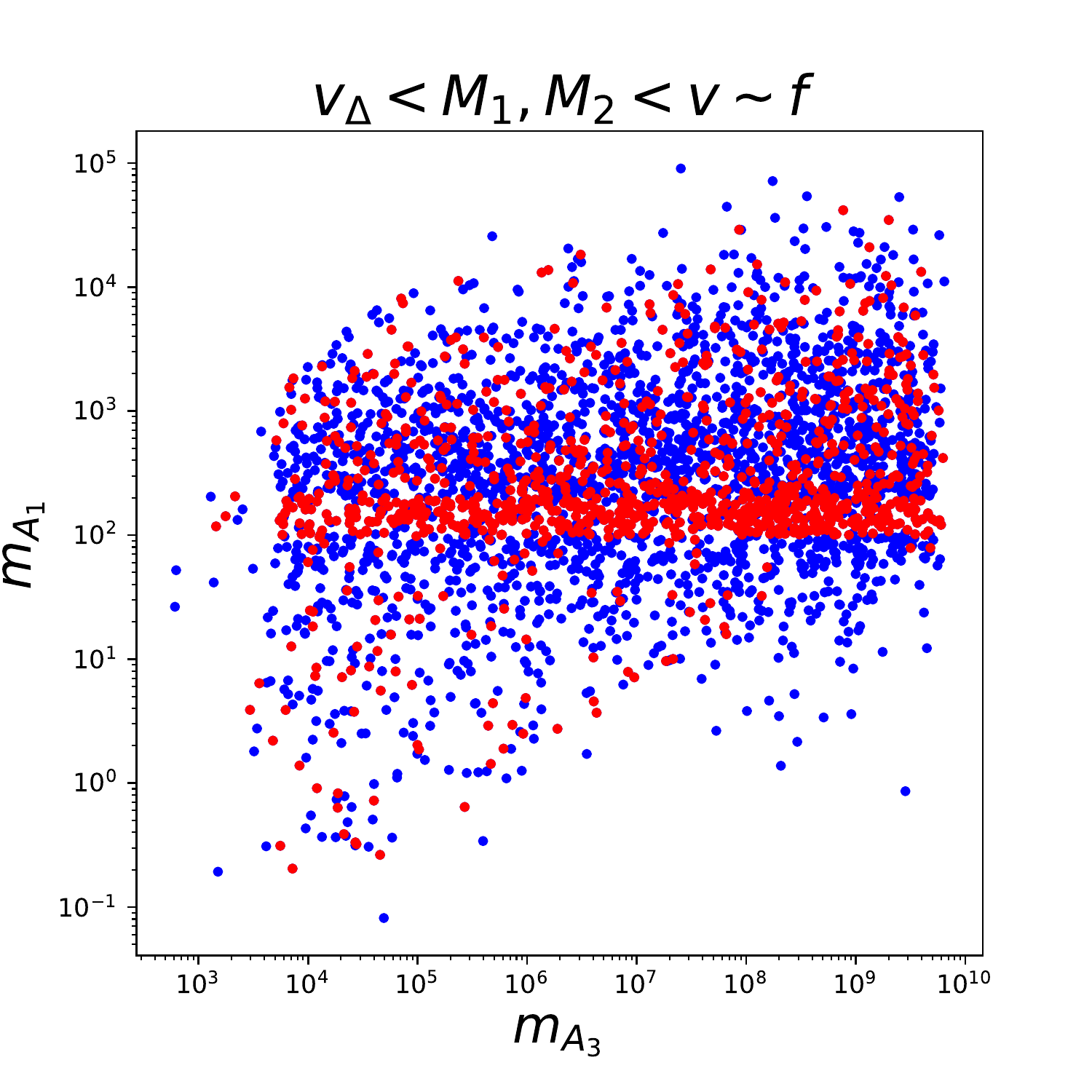}   
    \label{fig4b:b} 
    \vspace{4ex}
  \end{subfigure} 
 \caption{CP odd at $M_1,M_2<f\sim v$. The blue dots represents all possible positive definite masses of the scalars at tree-level and the red dots represents the positive definite masses of the scalars with the Higgs-like particle mass ($m_{h_1}$) lying between $100$ GeV and $200$ GeV(See the text).}
  \label{fig4b} 
\end{figure}

The CP-odd sector have three massive particles. Among them, there is one massive particle, $A_1$, that is a composition between $I_\rho$ and $I_\eta$, and its mass depends on the value of $f$. Since we are assuming that $f$ is varying in a EWSB energy scale, it was expected that such particle would have a mass in a EWSB energy scale too.
The second particle ($A_2$) is dominantly composed by $I_\Delta$ with its mass proportional to ratio between the scale of the two explicitly lepton number breaking parameters $M_1$ and $M_2$ and the scale of the spontaneous lepton number breaking parameter $v_\Delta$ and the scale of $v$. Since we are assuming that $v_\Delta< M_1,\;M_2$, the mass of the pseudoscalar $A_2$ is always equal or larger than the scale of the EWSB energy scale. At first, these two particles are very problematic for a low mass regime, since they can induce a decay of the Higgs $h_1$ into two of these pseudoscalars and the invisible decay of the $Z$ into three pseudoscalars. However, one interesting fact is that, imposing the correct mass of the Higgs ($h_1 \sim 125$ GeV), these two pseudoscalars naturally tends to be in the same mass scale of $h_1$, as can be seem in FIG \ref{fig4b}.

\begin{figure}[ht] 
  \begin{subfigure}[b]{0.5\linewidth}
    \centering    \includegraphics[width=0.75\linewidth]{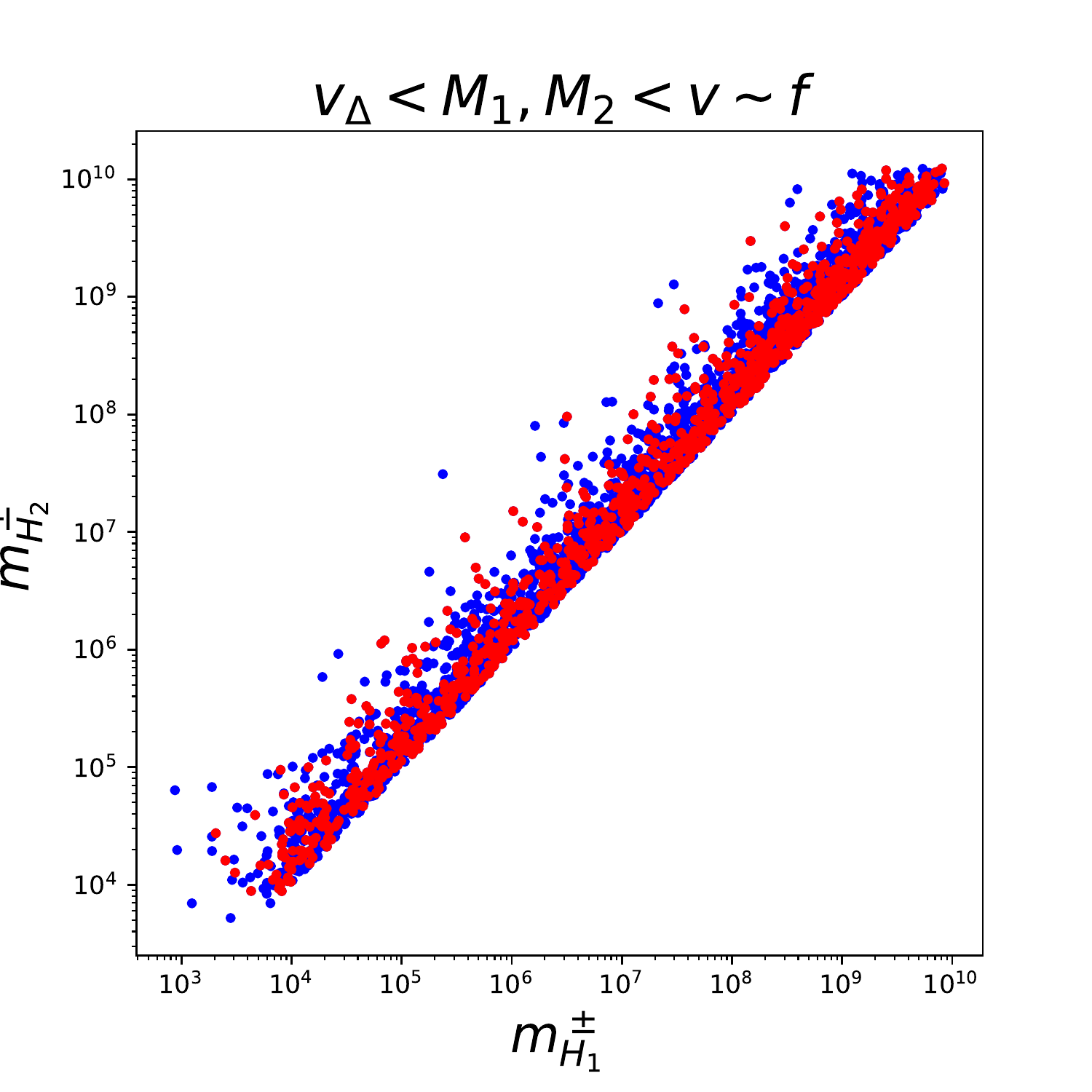}    \caption{331 singly-charged scalars}    
    \label{fig7b:a} 
    \vspace{4ex}
  \end{subfigure}
  \begin{subfigure}[b]{0.5\linewidth}
    \centering
\includegraphics[width=0.75\linewidth]{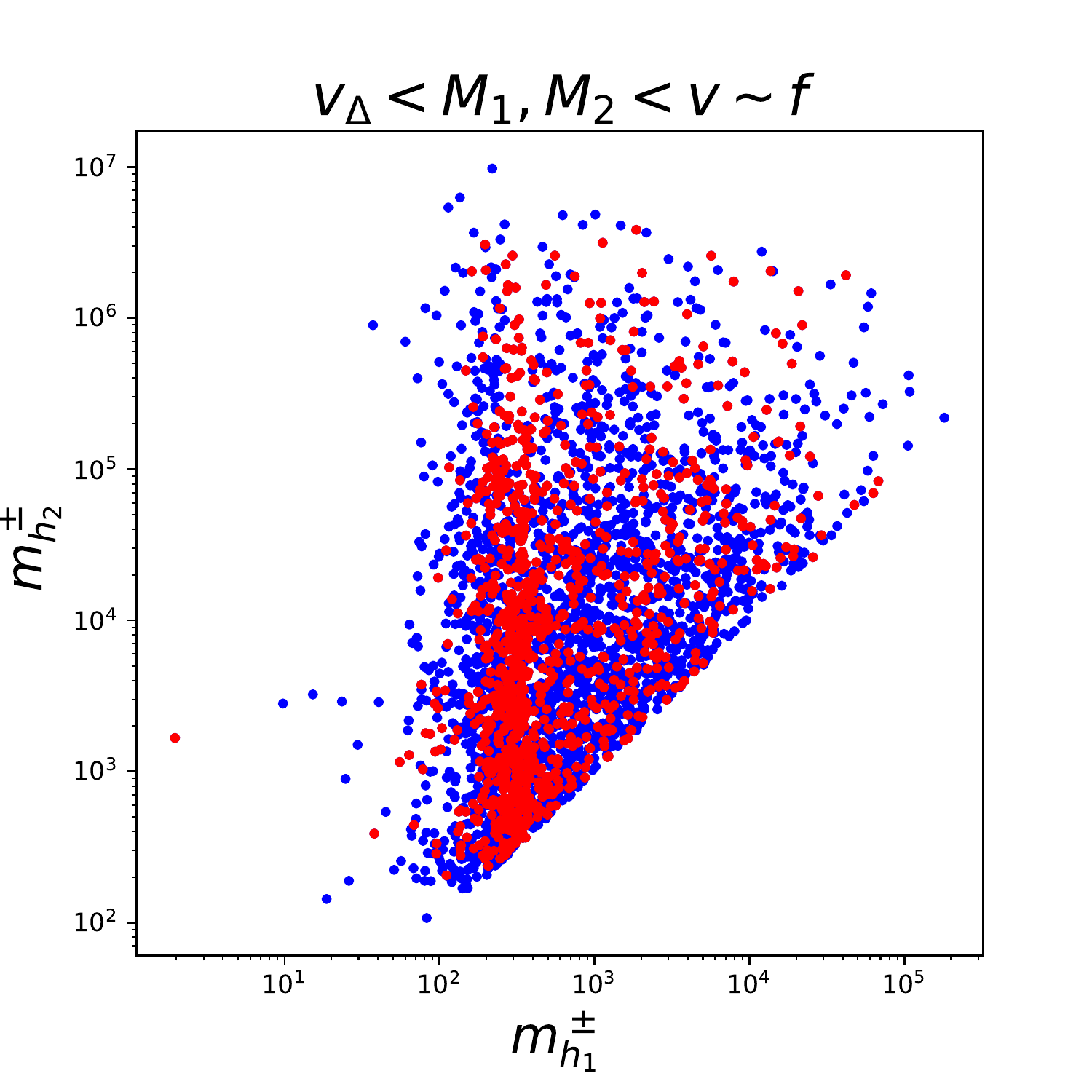}   \caption{EW singly-charged scalars} 
    \label{fig7b:b} 
    \vspace{4ex}
  \end{subfigure} 
 \caption{Singly charged scalars at $M_1,M_2<f\sim v$. The blue dots represents all possible positive definite masses of the scalars at tree-level and the red dots represents the positive definite masses of the scalars with the Higgs-like particle mass ($m_{h_1}$) lying between $100$ GeV and $200$ GeV(See the text). }
  \label{fig7b} 
\end{figure}

The singly-charged sector, as discussed previously, is composed by four massive particles (considering a particle and its antiparticle as one), two of them are typically from the EWSB scale, as can be seen in FIG. \ref{fig7b:b}. Among them, the first (${h_1}^\pm$) is a composition of $\eta^\pm$ and $\rho^\pm$ and the other (${h_2}^\pm$) is composed basically by $\Delta^\pm$. Imposing the correct Higgs mass, these charged scalars are typically heavier than $m_{h_1}$ (there are some points in which ${h_1}^\pm$ is allowed to have lower masses, however it is not a rule, but an exception). 

A comment about the doubly-charged particle in FIG. \ref{fig7b}. Such scalar is almost degenerated with the scalar ${h_2}^\pm$, since the last is predominantly $\Delta^\pm$. However, such scalar could be very light (tens of GeV), even for a correct Higgs mass, differently from the other scalars dominantly $\Delta$.

\begin{figure}[ht] 
    \centering    \includegraphics[width=0.4\linewidth]{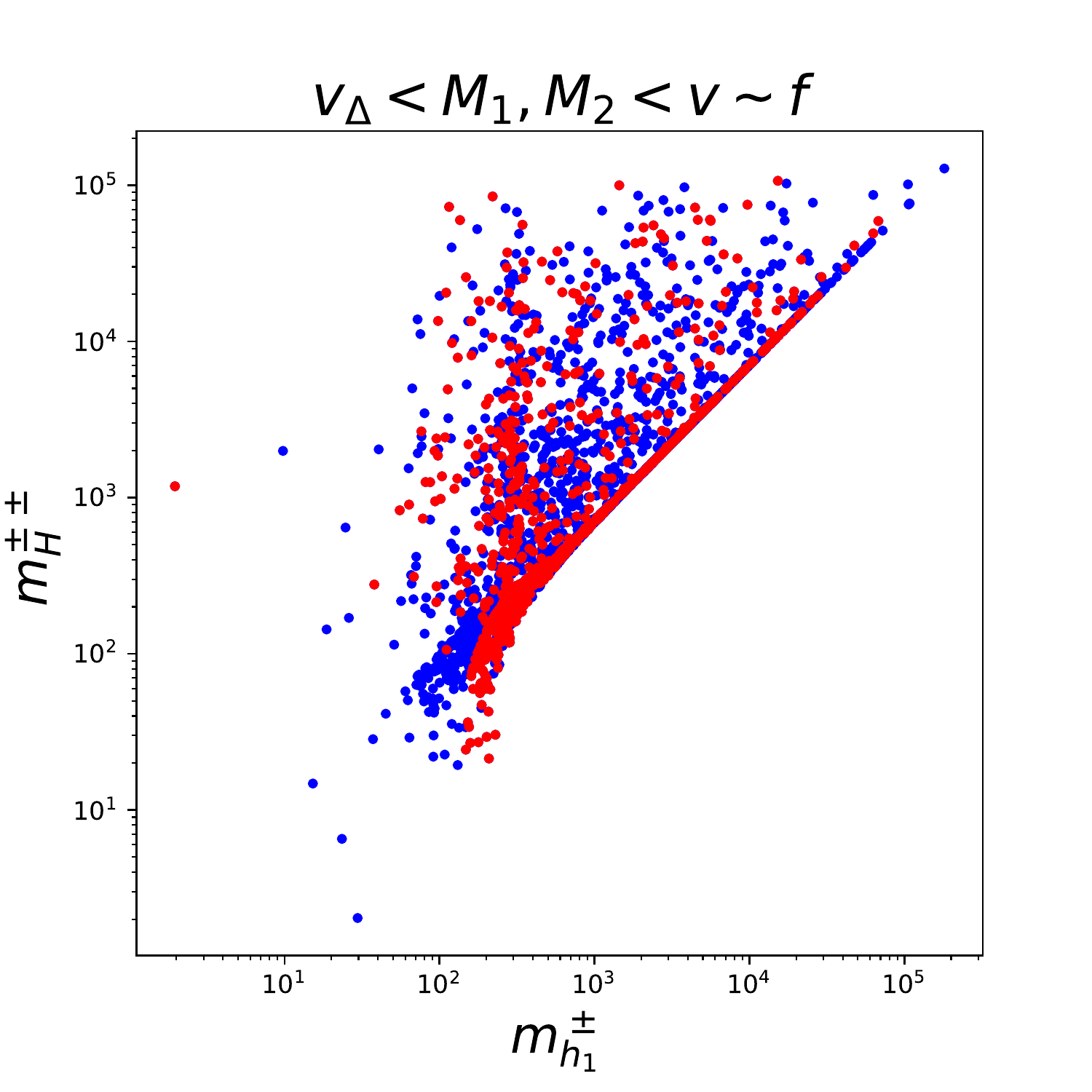}    \caption{Doubly charged scalar at $M_1,M_2<f\sim v$  .  The blue dots represents all possible positive definite masses of the scalars at tree-level and the red dots represents the positive definite masses of the scalars with the Higgs-like particle mass ($m_{h_1}$) lying between $100$ GeV and $200$ GeV(See the text).}    
    \vspace{4ex}
  \label{fig10b} 
\end{figure}

\begin{figure}[ht] 
    \centering    \includegraphics[width=0.4\linewidth]{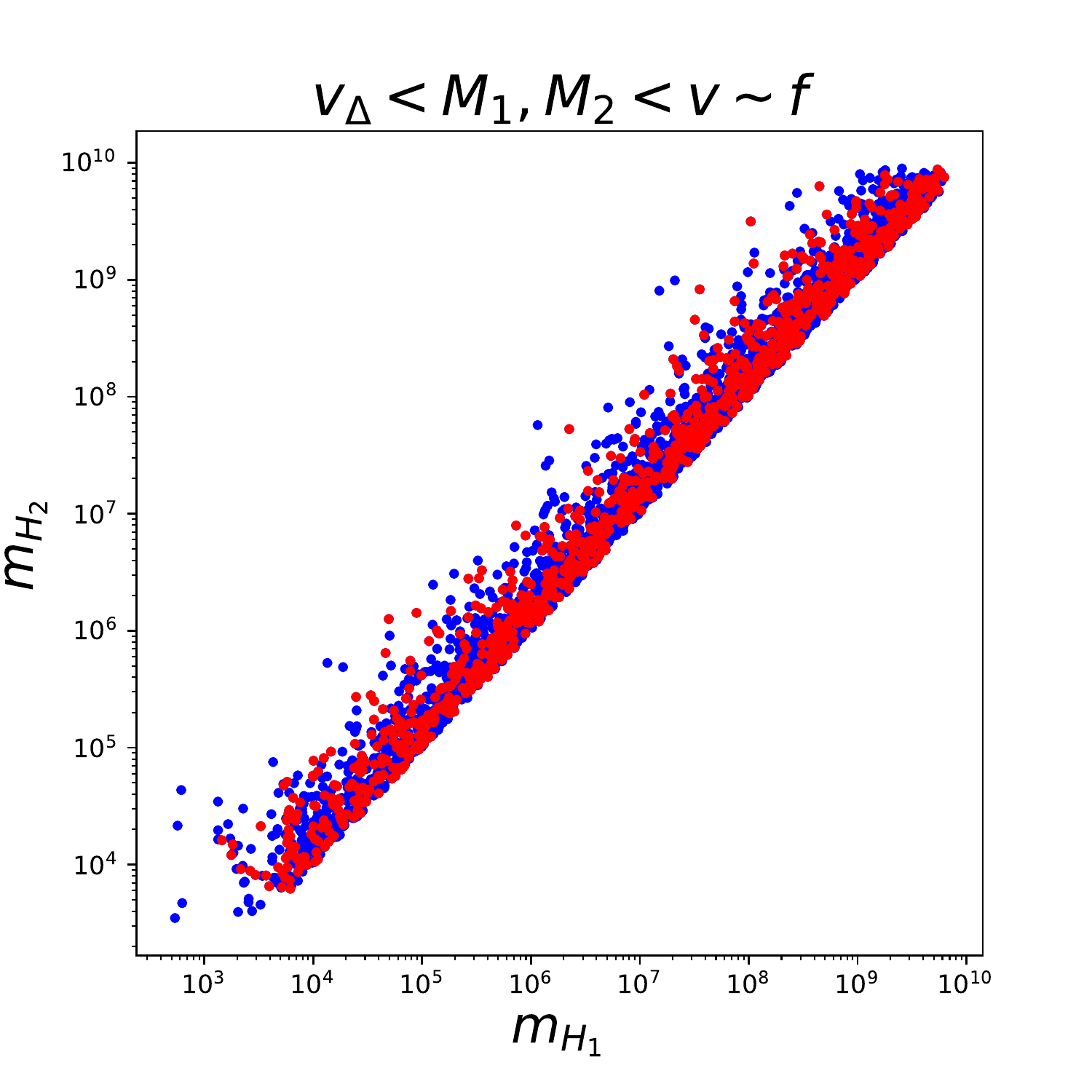}    \caption{Complex neutral scalar at $M_1,M_2<f\sim v$.  The blue dots represents all possible positive definite masses of the scalars at tree-level and the red dots represents the positive definite masses of the scalars with the Higgs-like particle mass ($m_{h_1}$) lying between $100$ GeV and $200$ GeV(See the text).}    
    \vspace{4ex}
  \label{fig13b} 
\end{figure}

\clearpage

\subsubsection{$M_1,M_2 \sim f\sim v$}

\begin{figure}[ht] 
  \begin{subfigure}[b]{0.5\linewidth}
    \centering\includegraphics[width=0.75\linewidth]{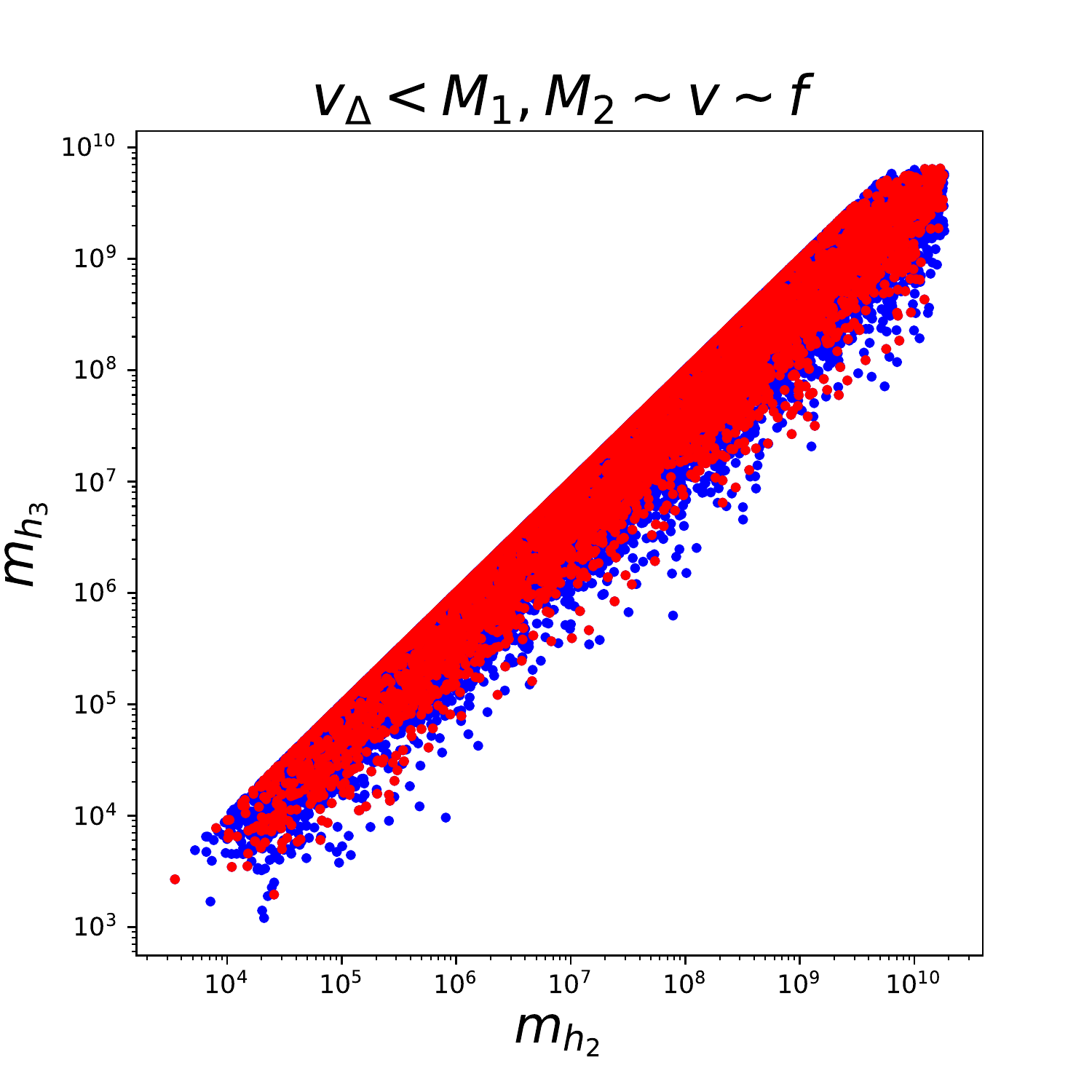} 
    \label{fig2b:a} 
    \vspace{4ex}
  \end{subfigure}
  \begin{subfigure}[b]{0.5\linewidth}
    \centering
\includegraphics[width=0.75\linewidth]{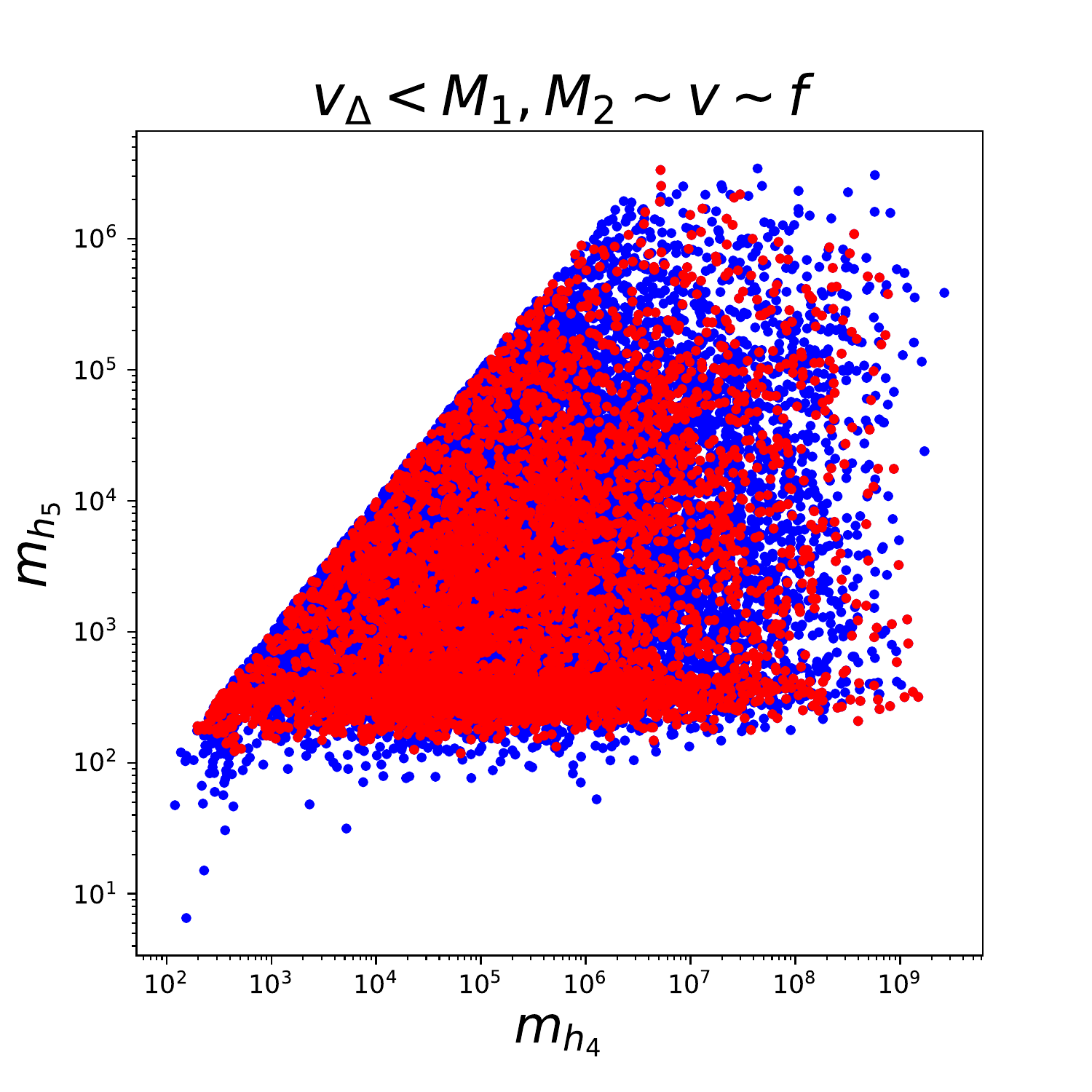} 
    \label{fig2b:b} 
    \vspace{4ex}
  \end{subfigure} 
  \begin{subfigure}[b]{0.5\linewidth}
    \centering    \includegraphics[width=0.75\linewidth]{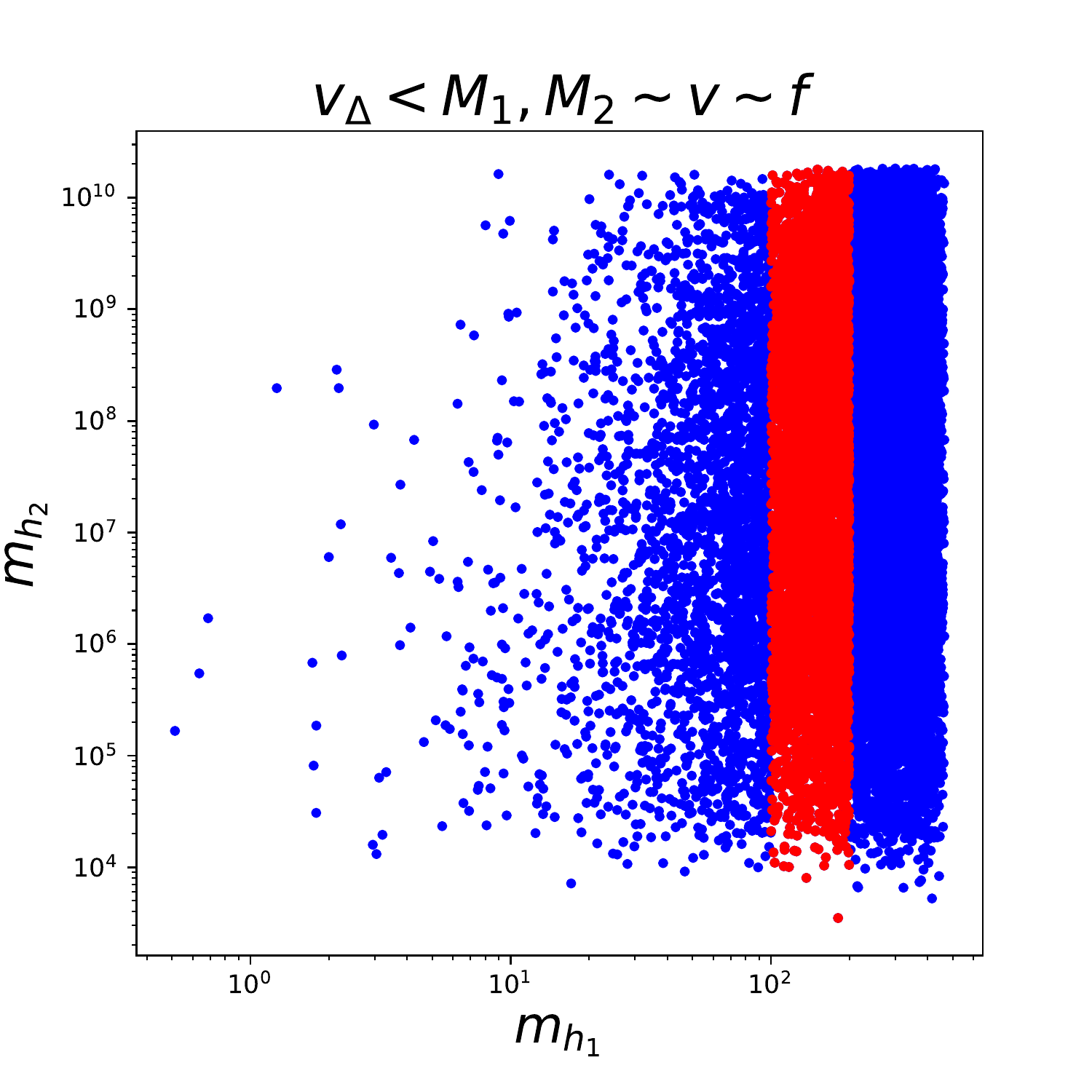}
    \label{fig2b:c} 
  \end{subfigure}
\caption{CP even at $M_1,M_2\sim f\sim v$. The blue dots represents all possible positive definite masses of the scalars at tree-level and the red dots represents the positive definite masses of the scalars with the Higgs-like particle mass ($m_{h_1}$) lying between $100$ GeV and $200$ GeV(See the text).}
  \label{fig2b} 
\end{figure}

In the CP-even sector, as can be seem in FIG. \ref{fig2b}, we observe the same pattern of the previous case. Among the five particles mentioned above, two of them, again, have masses at the the EWSB energy scale ($h_1$ and $h_5$) and they  are a perpendicular composition of $R_\eta$ and $R_\rho$. Here again $h_4$  is almost predominantly $R_\Delta$. Imposing that $M_1$ and $M_2$ are in a EWSB energy scale did not alter the CP-even sector, since the mixing between the $5\times 5$ matrix is so complex that somehow absorbed the effect of these parameters.

\begin{figure}[ht] 
  \begin{subfigure}[b]{0.5\linewidth}
    \centering    \includegraphics[width=0.75\linewidth]{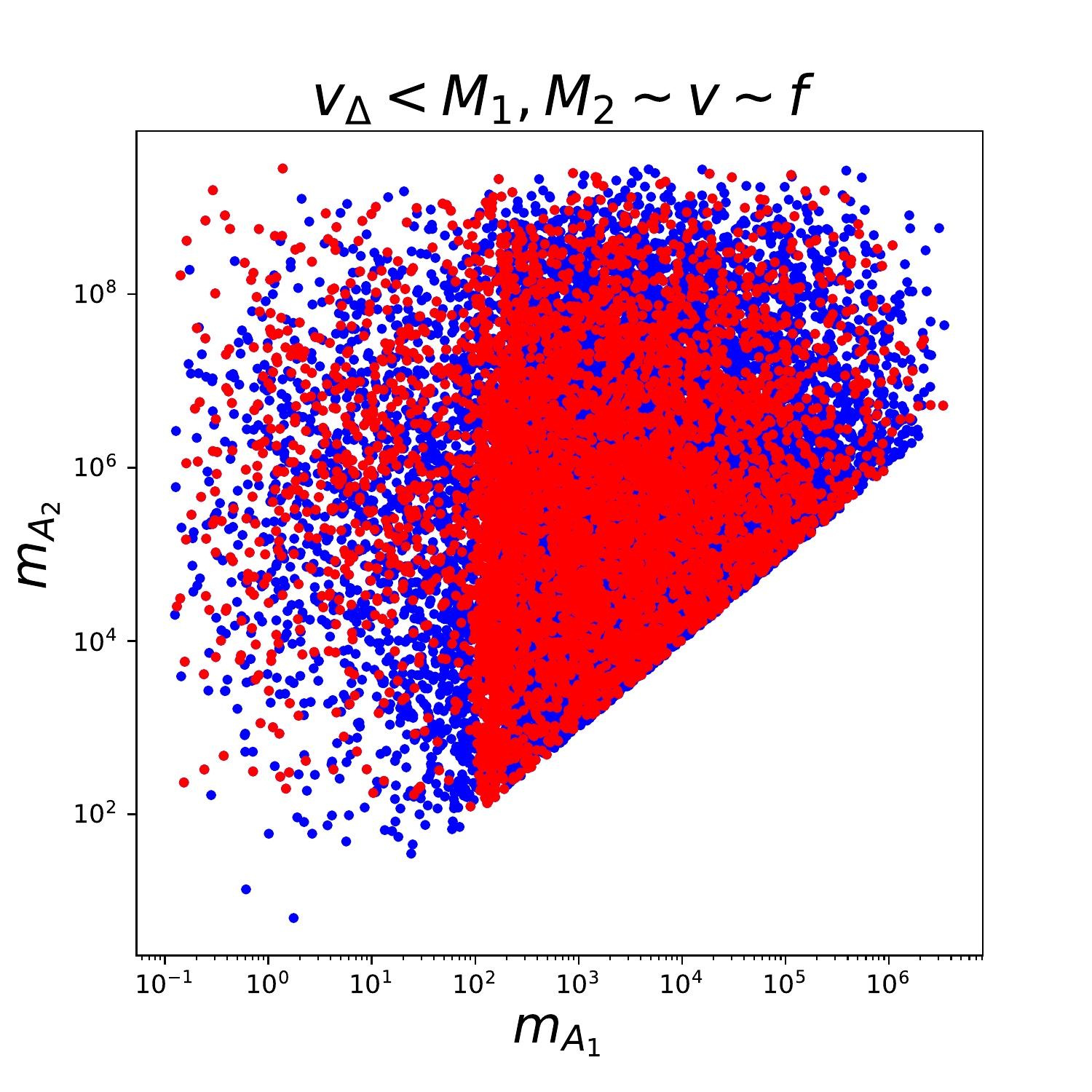}    
    \label{fig5b:a} 
    \vspace{4ex}
  \end{subfigure}
  \begin{subfigure}[b]{0.5\linewidth}
    \centering
\includegraphics[width=0.75\linewidth]{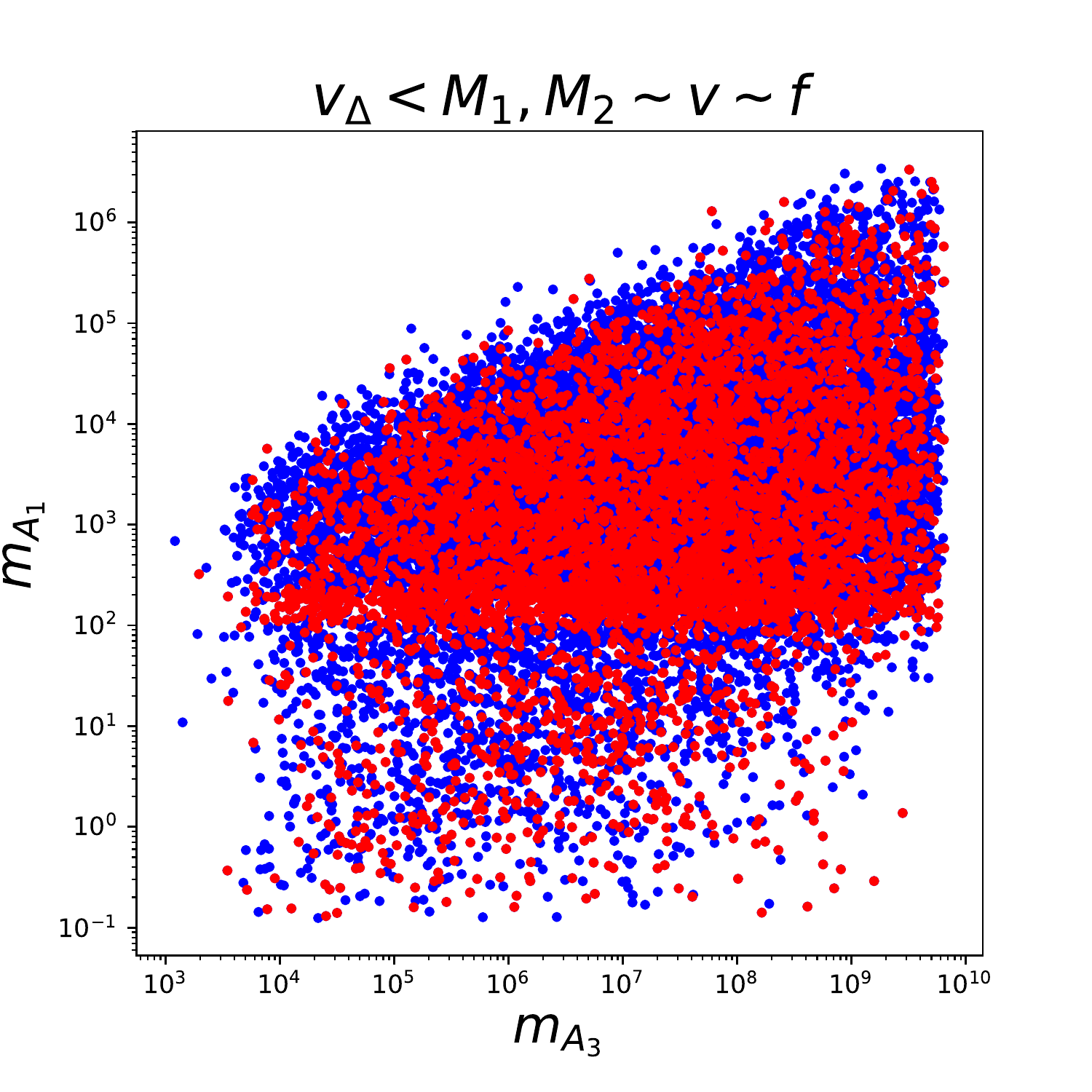}   
    \label{fig5b:b} 
    \vspace{4ex}
  \end{subfigure} 
 \caption{CP odd at $M_1,M_2\sim f\sim v$. The blue dots represents all possible positive definite masses of the scalars at tree-level and the red dots represents the positive definite masses of the scalars with the Higgs-like particle mass ($m_{h_1}$) lying between $100$ GeV and $200$ GeV(See the text).}
  \label{fig5b} 
\end{figure}

The CP-odd sector, as before, has one massive particle, $A_1$, that is a composition between $I_\rho$ and $I_\eta$, and its mass depends directly on the value of $f$. The particle $A_2$  is dominantly $I_\Delta$ and its mass is proportional to the scale of $M_1$ and $M_2$, and as in the CP-even case, the CP-odd case did not changed dramatically in comparison with the first case analysed, as can be seen in FIG. \ref{fig5b}.

\begin{figure}[ht] 
  \begin{subfigure}[b]{0.5\linewidth}
    \centering    \includegraphics[width=0.75\linewidth]{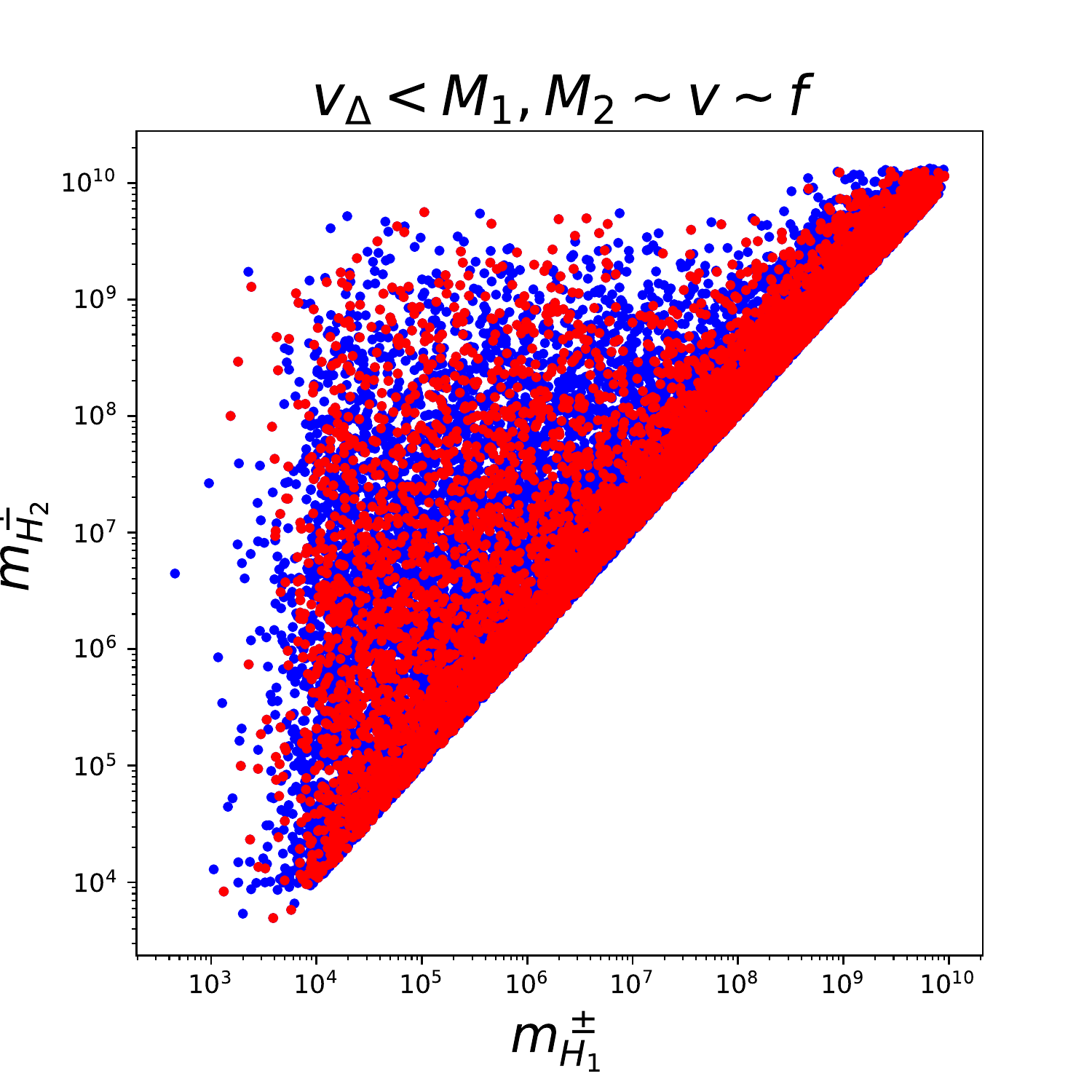}    \caption{331 singly-charged scalars}    
    \label{fig88:a} 
    \vspace{4ex}
  \end{subfigure}
  \begin{subfigure}[b]{0.5\linewidth}
    \centering
\includegraphics[width=0.75\linewidth]{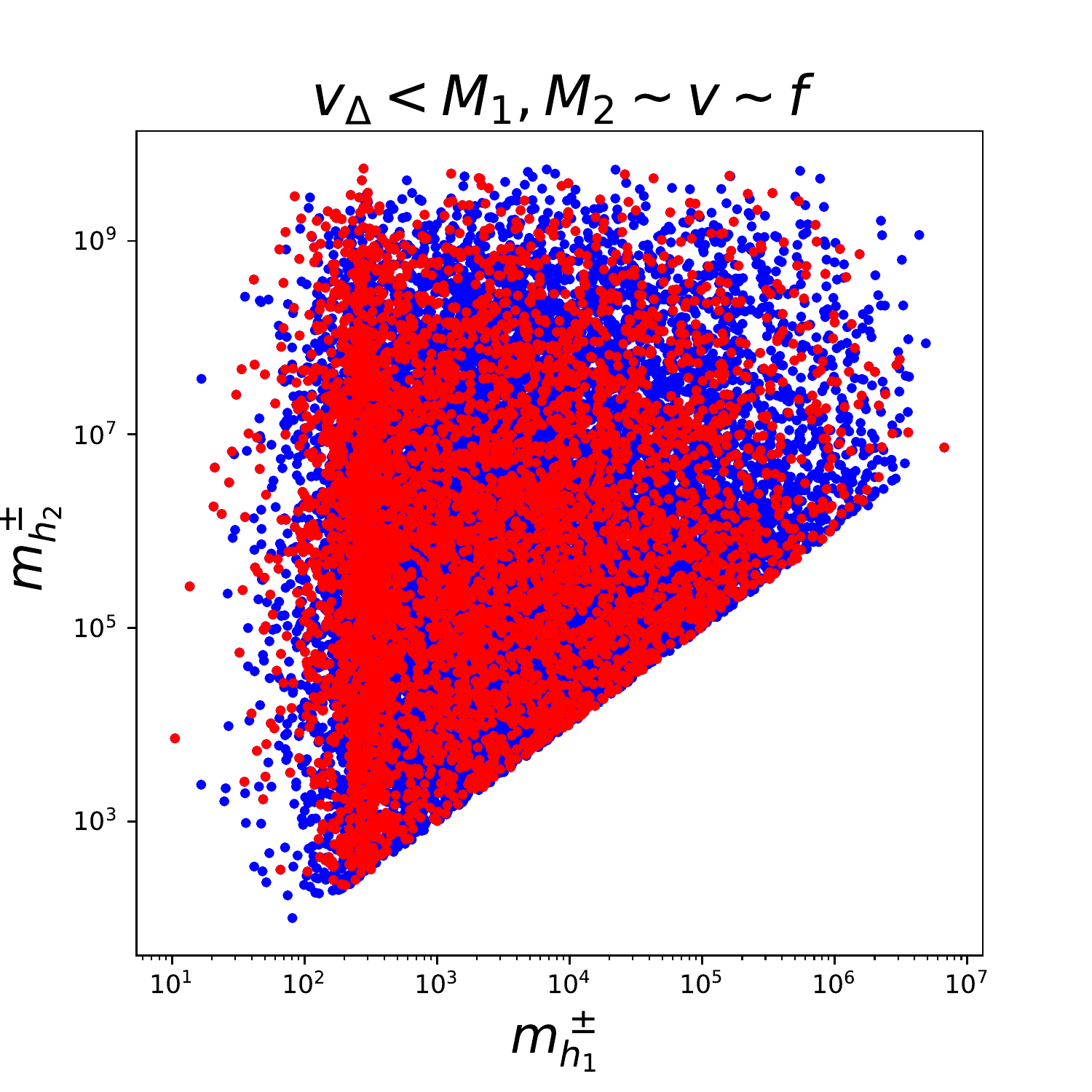}   
\caption{EW singly-charged scalars} 
    \label{fig88:b} 
    \vspace{4ex}
  \end{subfigure} 
 \caption{Singly charged scalars at $M_1,M_2\sim f\sim v$. The blue dots represents all possible positive definite masses of the scalars at tree-level and the red dots represents the positive definite masses of the scalars with the Higgs-like particle mass ($m_{h_1}$) lying between $100$ GeV and $200$ GeV(See the text). }
  \label{fig88} 
\end{figure}

As before, among the four massive singly-charged scalars, two of them are typically from the EWSB scale, as can be seen in FIG. \ref{fig88:b}. One (${h_1}^\pm$) is a composition of $\eta^\pm$ and $\rho^\pm$ and the other (${h_2}^\pm$) is composed basically by $\Delta^\pm$. Imposing the correct Higgs mass ($m_{h_1}\sim 125$ GeV), the mass of these charged scalars are typically higher than $m_{h_1}$ (there are some points in which ${h_1}^\pm$ is allowed to have lower masses, however it is not a rule, but an exception). 

\begin{figure}[ht] 
    \centering    \includegraphics[width=0.4\linewidth]{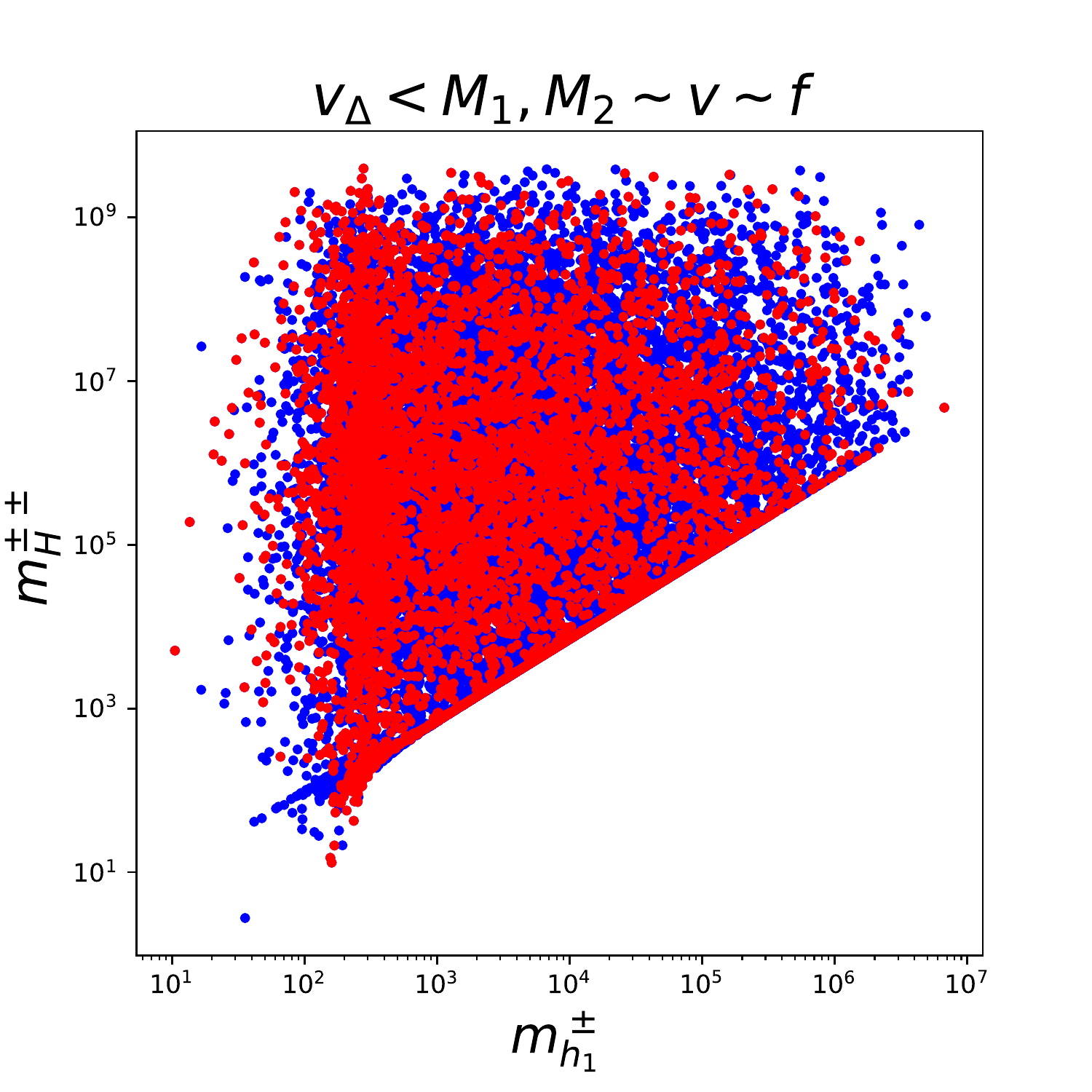}    \caption{$M_1,M_2 \sim f\sim v$  Doubly charged scalar. The blue dots represents all possible positive definite masses of the scalars at tree-level and the red dots represents the positive definite masses of the scalars with the Higgs-like particle mass ($m_{h_1}$) lying between $100$ GeV and $200$ GeV(See the text).}    
    \vspace{4ex}
  \label{fig11b} 
\end{figure}

\begin{figure}[ht] 
    \centering    \includegraphics[width=0.4\linewidth]{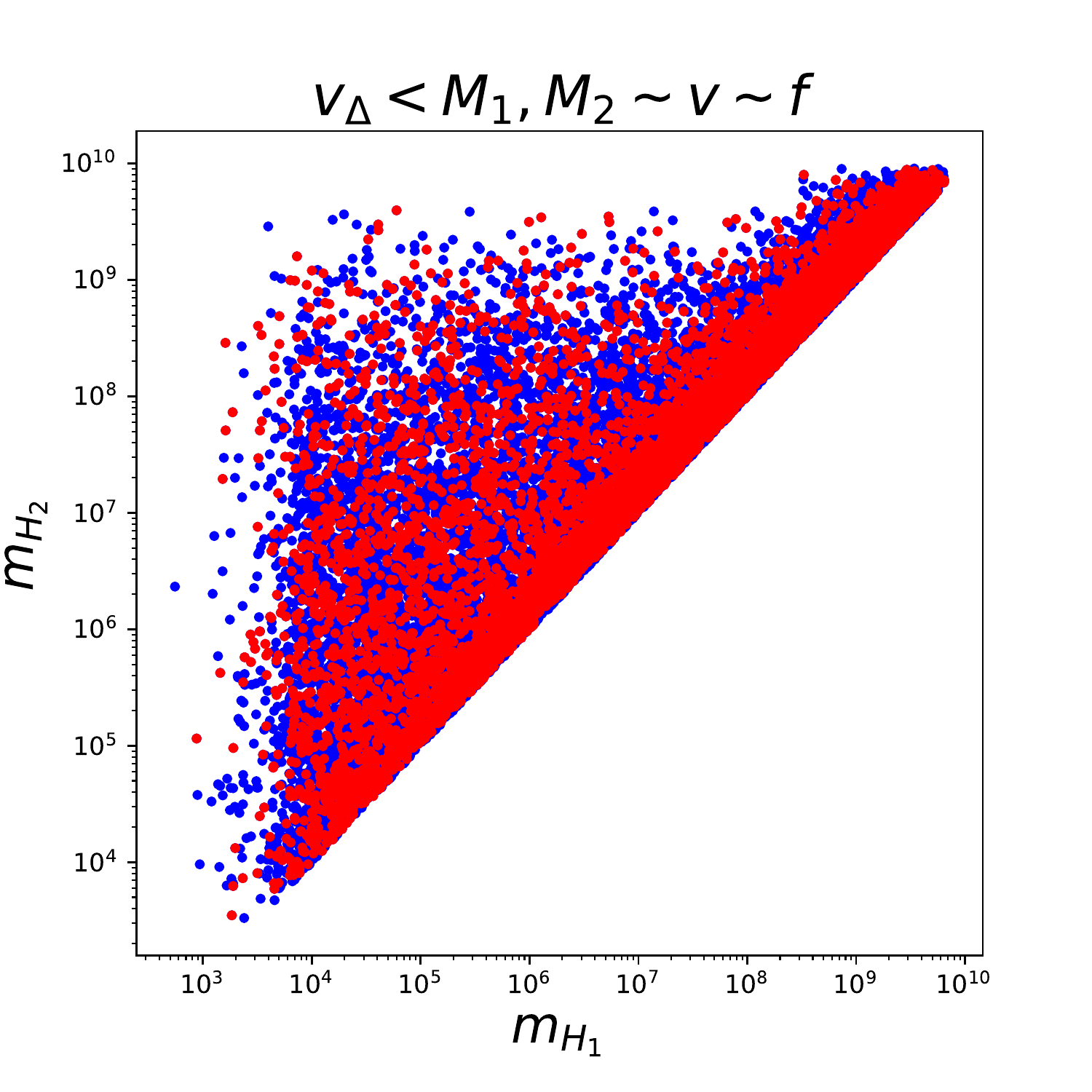}    \caption{Complex neutral scalar at$M_1,M_2\sim f\sim v$. The blue dots represents all possible positive definite masses of the scalars at tree-level and the red dots represents the positive definite masses of the scalars with the Higgs-like particle mass ($m_{h_1}$) lying between $100$ GeV and $200$ GeV(See the text).}    
    \vspace{4ex}
  \label{fig14b} 
\end{figure}

\clearpage

\subsection{$v_\sigma$ and $v_\Delta \sim v$}
For this case, we are letting $f$ to vary between $10^{-3}$ GeV and $10^{3}$ GeV. The vev's of the scalars $\eta$ and $\rho$ as  $v_\rho = \sqrt{246.11^2\;\mathrm{GeV}^2-v_\eta^2}$ and  $0$ GeV$<v_\eta<246.11$ GeV and
the vev of the triplet $\chi$ is varying between $10^3\;\mathrm{GeV}<v_{\chi^\prime}<10^{10}$ GeV. The vev of the triplet of the sextet ($v_\Delta$) is varying as $10^{-4}$ GeV$<v_\Delta <1$ GeV and the vev of the singlet of the sextet ($v_\sigma$) is varying between 
$10^{-4}$ GeV$<v_\sigma <10^{3}$ GeV. They always respect the hierarchy among the vev's as $v_\Delta , v_\sigma<v_{\chi^\prime}$. Differently from before, we are not considering that $v_\Delta < M_1,\;M_2$. Hence, we are not interested in imposing a hierarchy among the lepton number symmetry breaking energy scales. To finish, as before, all quartic couplings are varying between $-\sqrt{4\pi}$ and $\sqrt{4\pi}$. All these parameters will vary between this interval no matter the energy regime of $M_1$ and $M_2$.

Here, we are varying the parameters $M_1$ and $M_2$ between two different energy regime, $M_1,M_2 < f$ and $M_1,M_2\sim f$. In the first case we are varying the parameters $M_1,\;M_2$ as  $10^{-12}\;\mathrm{GeV}<M_1,M_2<10^{-6}\;\mathrm{GeV}$ and in the second case we are varying the parameters $M_1,\;M_2$ as  $10^{-3}\;\mathrm{GeV}<M_1,M_2<10^{3}\;\mathrm{GeV}$.

\subsubsection{$  M_1,M_2 < f\sim v$}
Here, we recover the same physics of the THDM+triplet as before. The components of the doublet $\Phi$ and the singlet $\sigma$ decouples from the typical particles of the EWSB energy scale ($\rho$ and $\;\eta$ ) and they becomes exclusively related with the 331 typical energy scale, $v_{\chi^\prime}$. However, the components of the triplet $\Delta$ becomes very light. For example, the CP-odd particle $A_2$, dominantly $I_\Delta$, acquires a mass proportional to the ratio $\sqrt{(M/v_\Delta)v^2}$, and for $M\ll v_\Delta$, the mass of such pseudo-Nambu Goldstone becomes very light, as indicated in FIG. \ref{fig60b}.  
This means that, for such low masses, there is a high probability that the invisible decay of the $Z$ boson into three $A_2$ and the invisible decay of the $h_1$ into two $A_2$ becomes very high, excluded by actual experimental bounds.

\begin{figure}[ht] 
  \begin{subfigure}[b]{0.5\linewidth}
    \centering\includegraphics[width=0.75\linewidth]{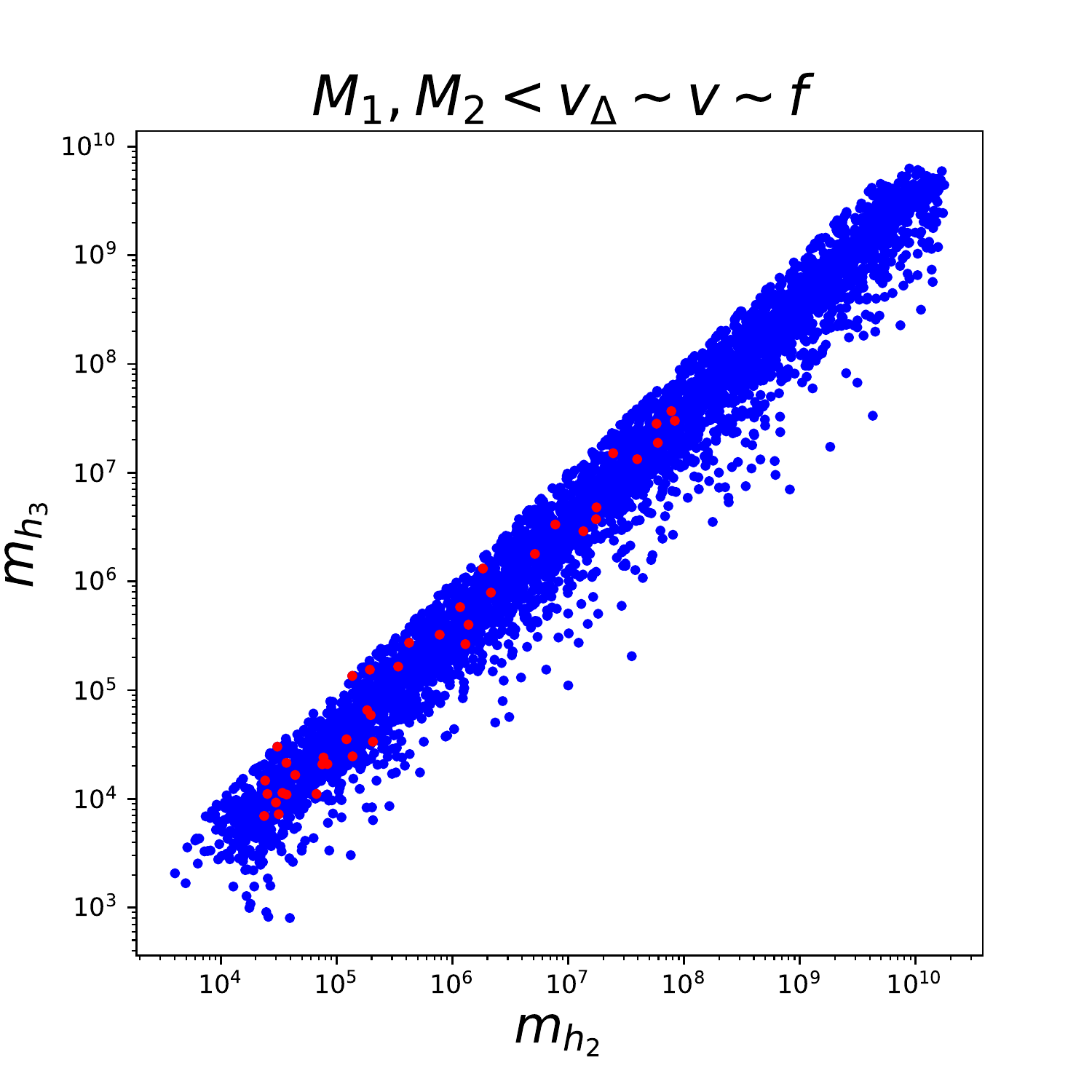} 
    \label{fig30b:a} 
    \vspace{4ex}
  \end{subfigure}
  \begin{subfigure}[b]{0.5\linewidth}
    \centering
\includegraphics[width=0.75\linewidth]{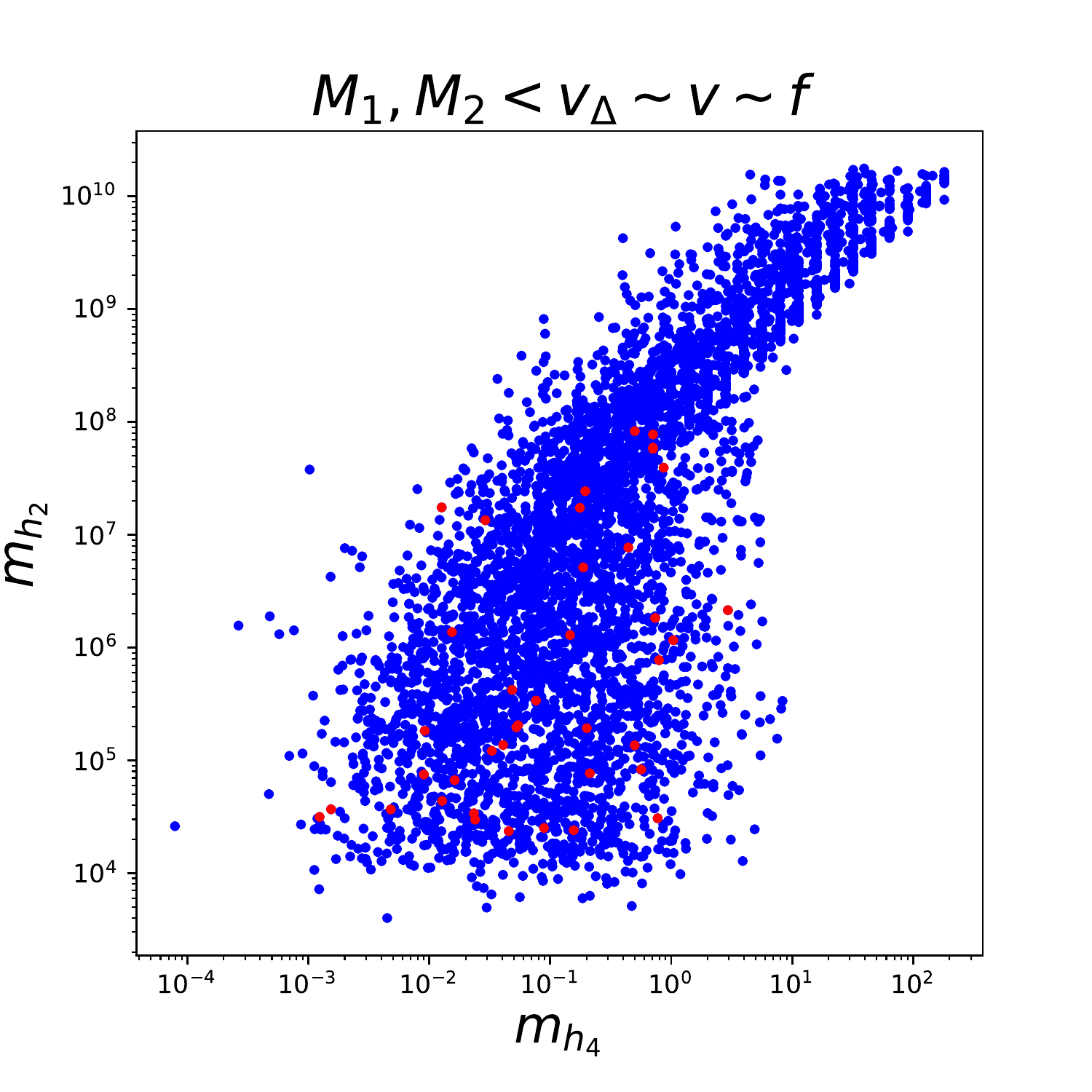} 
    \label{fig30b:b} 
    \vspace{4ex}
  \end{subfigure} 
  \begin{subfigure}[b]{0.5\linewidth}
    \centering    \includegraphics[width=0.75\linewidth]{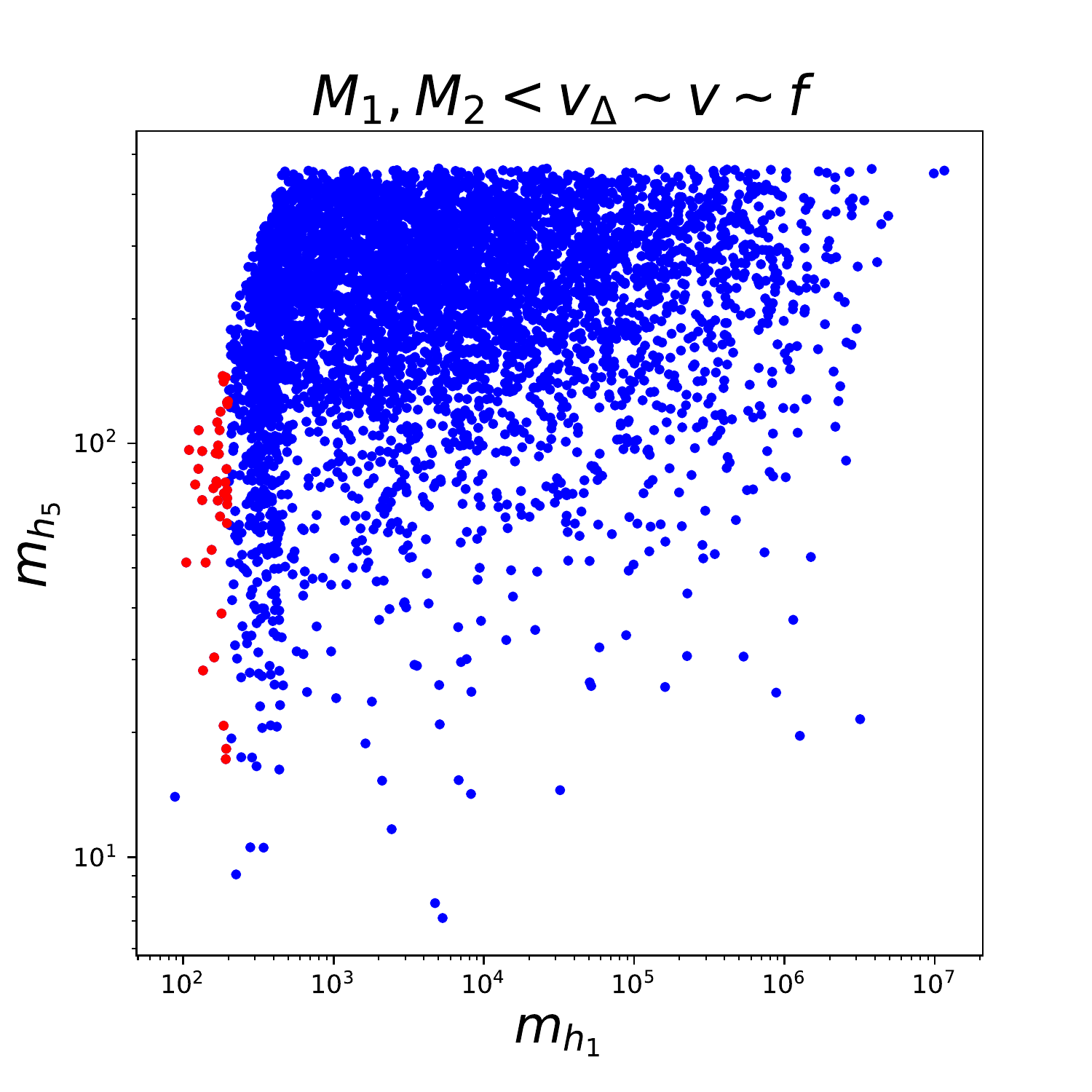}
    \label{fig30b:c} 
  \end{subfigure}
\caption{CP even at $  M_1,M_2 < v_\Delta  \sim f\sim v$ . The blue dots represents all possible positive definite masses of the scalars at tree-level and the red dots represents the positive definite masses of the scalars with the Higgs-like particle mass ($m_{h_1}$) lying between $100$ GeV and $200$ GeV(See the text).}
  \label{fig30b} 
\end{figure}

\begin{figure}[ht] 
  \begin{subfigure}[b]{0.5\linewidth}
    \centering    \includegraphics[width=0.75\linewidth]{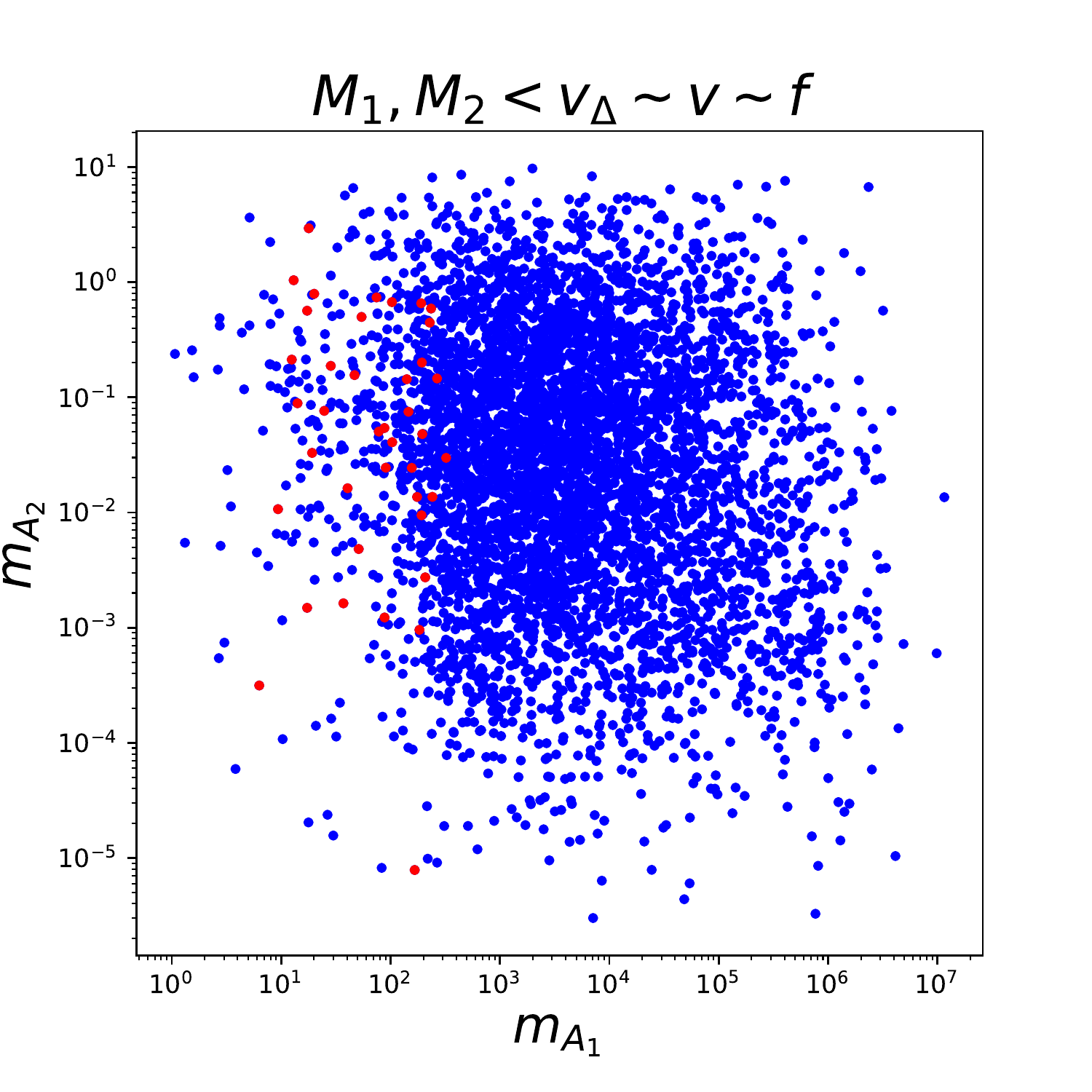}    
    \label{fig60b:a} 
    \vspace{4ex}
  \end{subfigure}
  \begin{subfigure}[b]{0.5\linewidth}
    \centering
\includegraphics[width=0.75\linewidth]{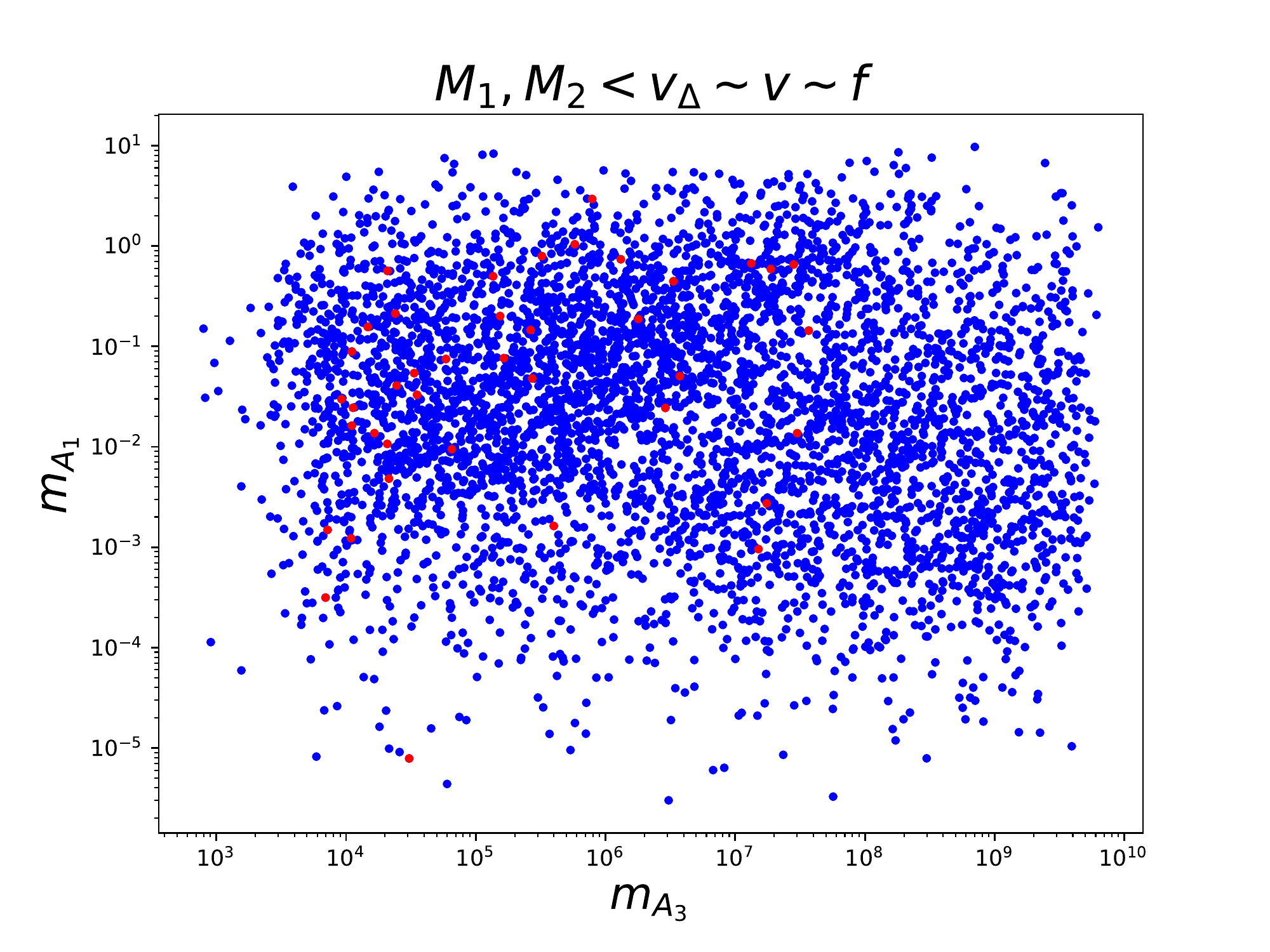}   
    \label{fig60b:b} 
    \vspace{4ex}
  \end{subfigure} 
 \caption{CP odd at $  M_1,M_2 < v_\Delta  \sim f\sim v$ . The blue dots represents all possible positive definite masses of the scalars at tree-level and the red dots represents the positive definite masses of the scalars with the Higgs-like particle mass ($m_{h_1}$) lying between $100$ GeV and $200$ GeV(See the text). }
  \label{fig60b} 
\end{figure}

\begin{figure}[ht] 
  \begin{subfigure}[b]{0.5\linewidth}
    \centering    \includegraphics[width=0.75\linewidth]{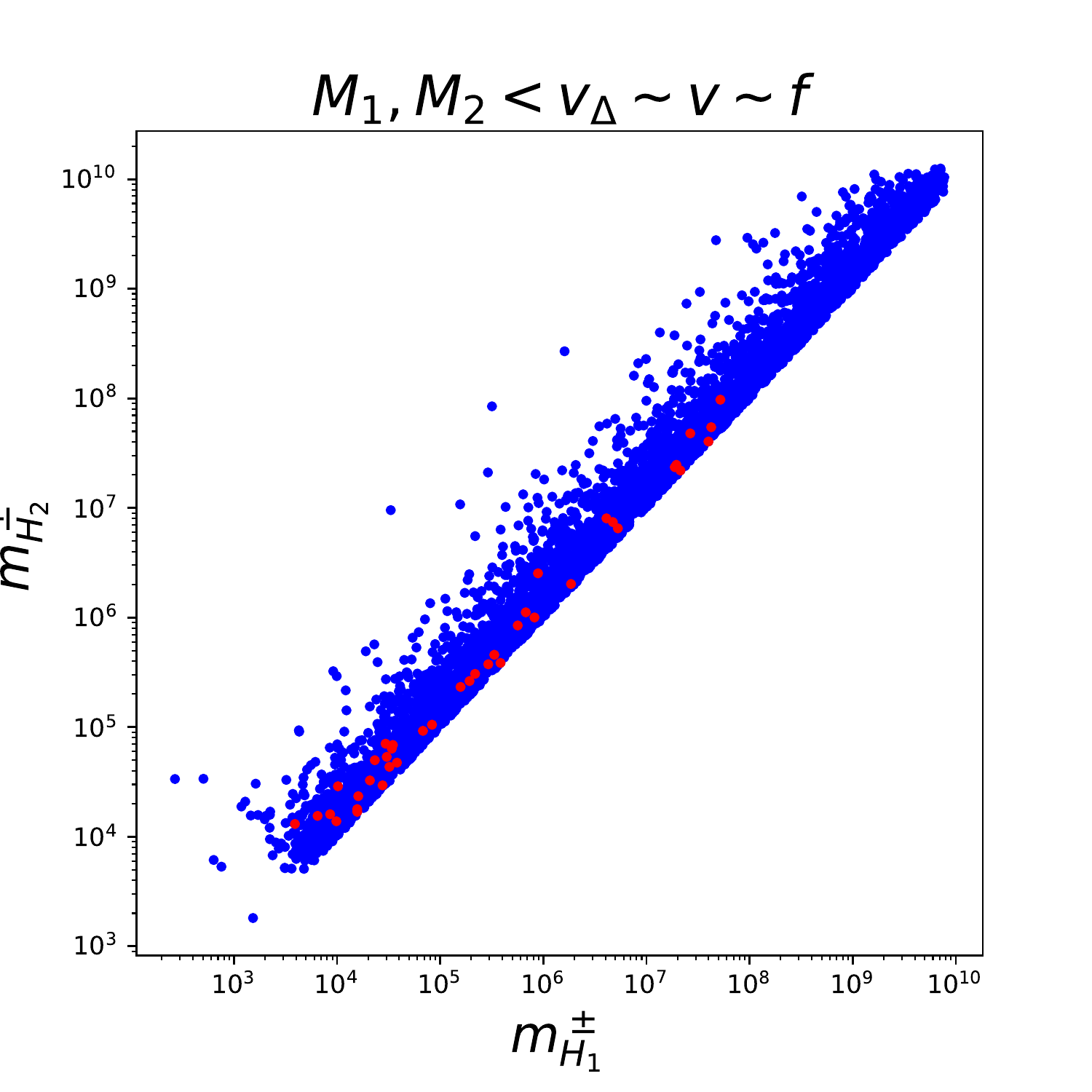}    
\caption{331 singly-charged scalars} 
    \label{fig90b:a} 
    \vspace{4ex}
  \end{subfigure}
  \begin{subfigure}[b]{0.5\linewidth}
    \centering
\includegraphics[width=0.75\linewidth]{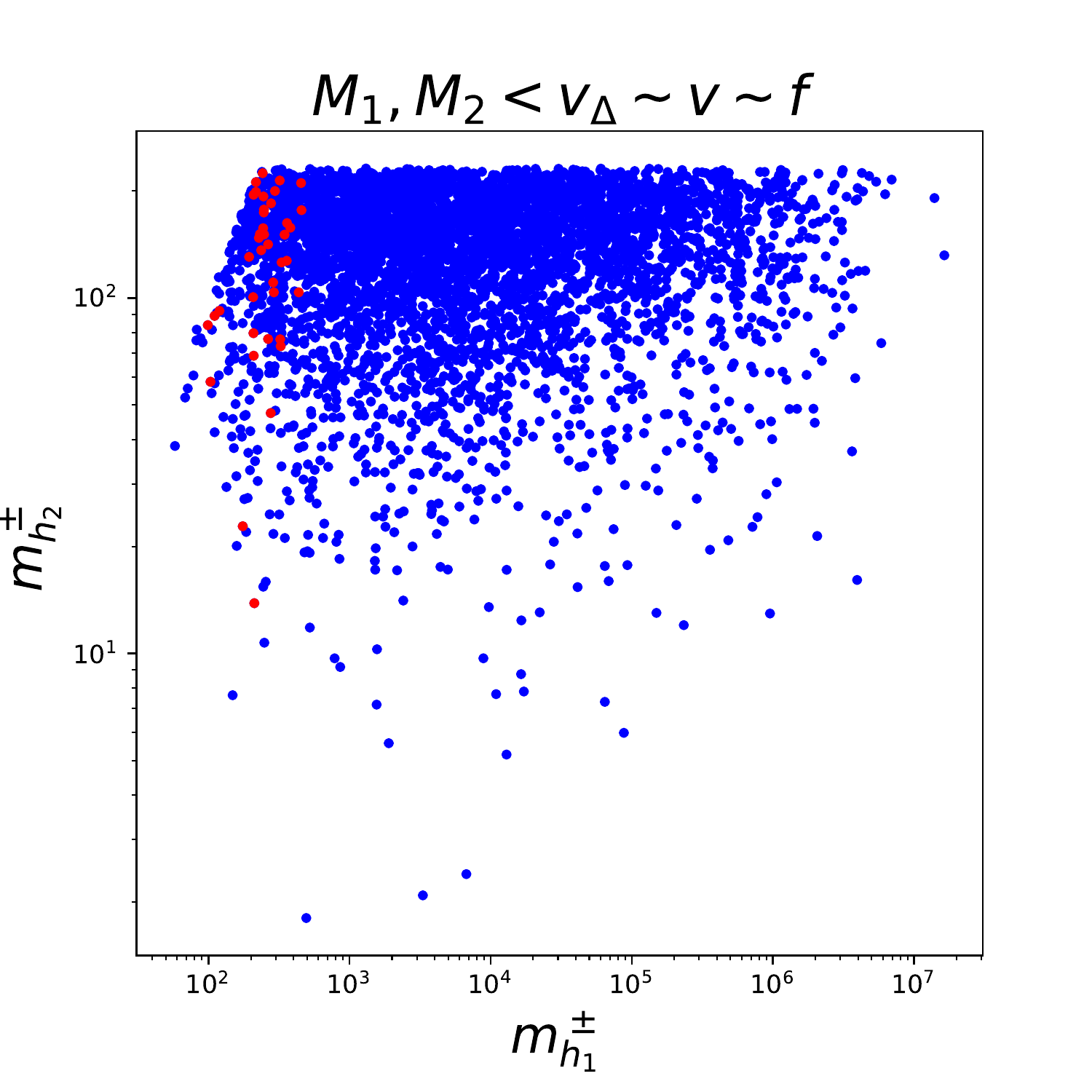}   
\caption{EW singly-charged scalars} 
    \label{fig90b:b} 
    \vspace{4ex}
  \end{subfigure} 
 \caption{Singly charged scalars at $  M_1,M_2 < v_\Delta  \sim f\sim v$ . The blue dots represents all possible positive definite masses of the scalars at tree-level and the red dots represents the positive definite masses of the scalars with the Higgs-like particle mass ($m_{h_1}$) lying between $100$ GeV and $200$ GeV(See the text).}
  \label{fig90b} 
\end{figure}

\begin{figure}[ht] 
    \centering    \includegraphics[width=0.4\linewidth]{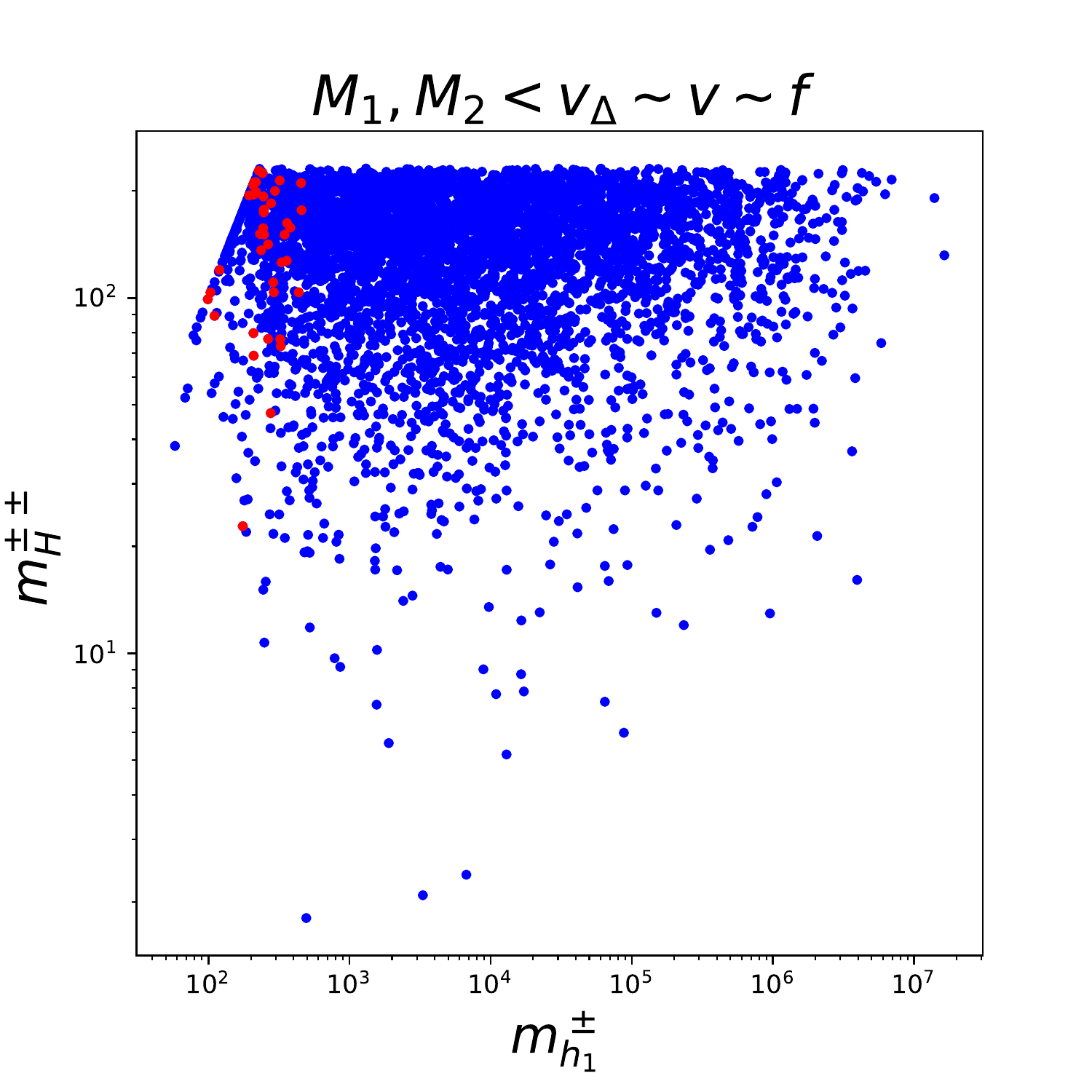}    \caption{Doubly charged scalar at $  M_1,M_2 < v_\Delta  \sim f\sim v$   . The blue dots represents all possible positive definite masses of the scalars at tree-level and the red dots represents the positive definite masses of the scalars with the Higgs-like particle mass ($m_{h_1}$) lying between $100$ GeV and $200$ GeV(See the text).}    
    \vspace{4ex}
  \label{fig120b} 
\end{figure}

The same explanation for the mass of the CP-even particle dominantly $R_\Delta$, $m_{h_4}$, represented in FIG. \ref{fig30b}.
The other particles dominantly composed by the components of the triplet $\Delta$ are $m_{H^{\pm\pm}}$ and $m_{h^{\pm}_2}$. In the FIGs. \ref{fig90b:b} and \ref{fig120b} we can see clearly that such particles are light. However, they lies in the typical EWSB energy scales. This means that, for natural values of the quartic couplings and for very low $M_1,\;M_2$, the mass of such scalars will depends on the vev's $v$ and $v_\Delta$. This means that, the mass of the charged scalars will always be bounded from below by the EWSB typical energy scale.

\begin{figure}[ht] 
    \centering    \includegraphics[width=0.4\linewidth]{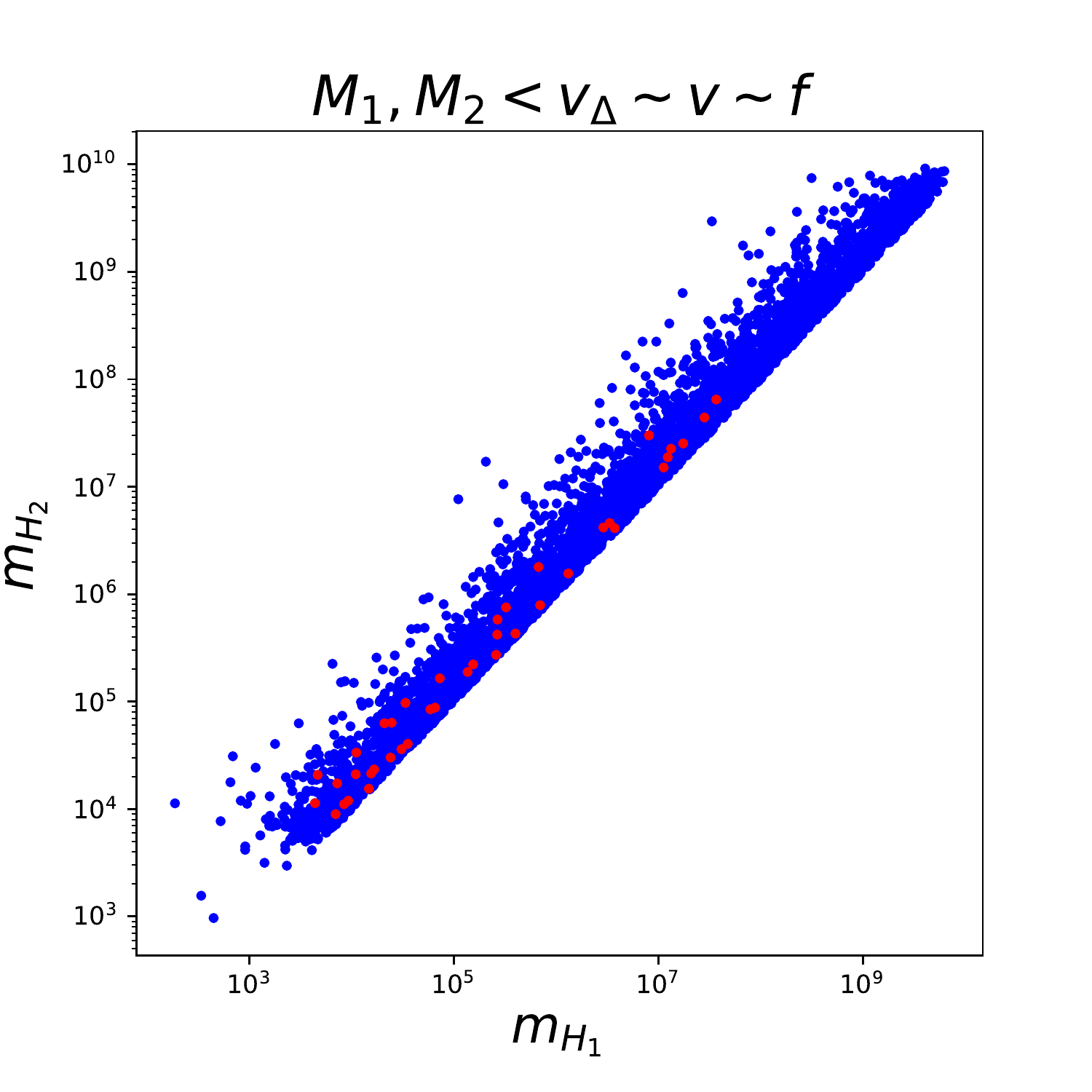}    \caption{Complex neutral scalar at $  M_1,M_2 < v_\Delta  \sim f\sim v$. The blue dots represents all possible positive definite masses of the scalars at tree-level and the red dots represents the positive definite masses of the scalars with the Higgs-like particle mass ($m_{h_1}$) lying between $100$ GeV and $200$ GeV(See the text).}    
    \vspace{4ex}
  \label{fig150b} 
\end{figure}

\clearpage

\subsubsection{$  M_1,M_2 \sim f\sim v$}
This case is the most phenomenologically rich among all. Here, we recover the same physics of the THDM+triplet as before, however, the components of the doublet $\Phi$ are allowed to become phenomenologically viable particles, too. The singly charged particle $H_1^\pm$ and the complex scalar particle $H_1$ are allowed to assume values between hundreds of GeV. However, this is true only for $v_{\chi^\prime}\sim 10^4 $ GeV. For higher values of such vev, the mass of the components of $\Phi$ becomes large enough to run away from the present experimental bounds.  
As the other cases, the particles of the triplet $\Delta$ can assume masses phenomenologically viable. However, for high values of $M$, the mass of the components of $\Delta$ tends to increase. One curious fact is the freedom of the $H^{\pm\pm}$ particle to have masses between 1 GeV and $10^2$ TeV.

\begin{figure}[ht] 
  \begin{subfigure}[b]{0.5\linewidth}
    \centering\includegraphics[width=0.75\linewidth]{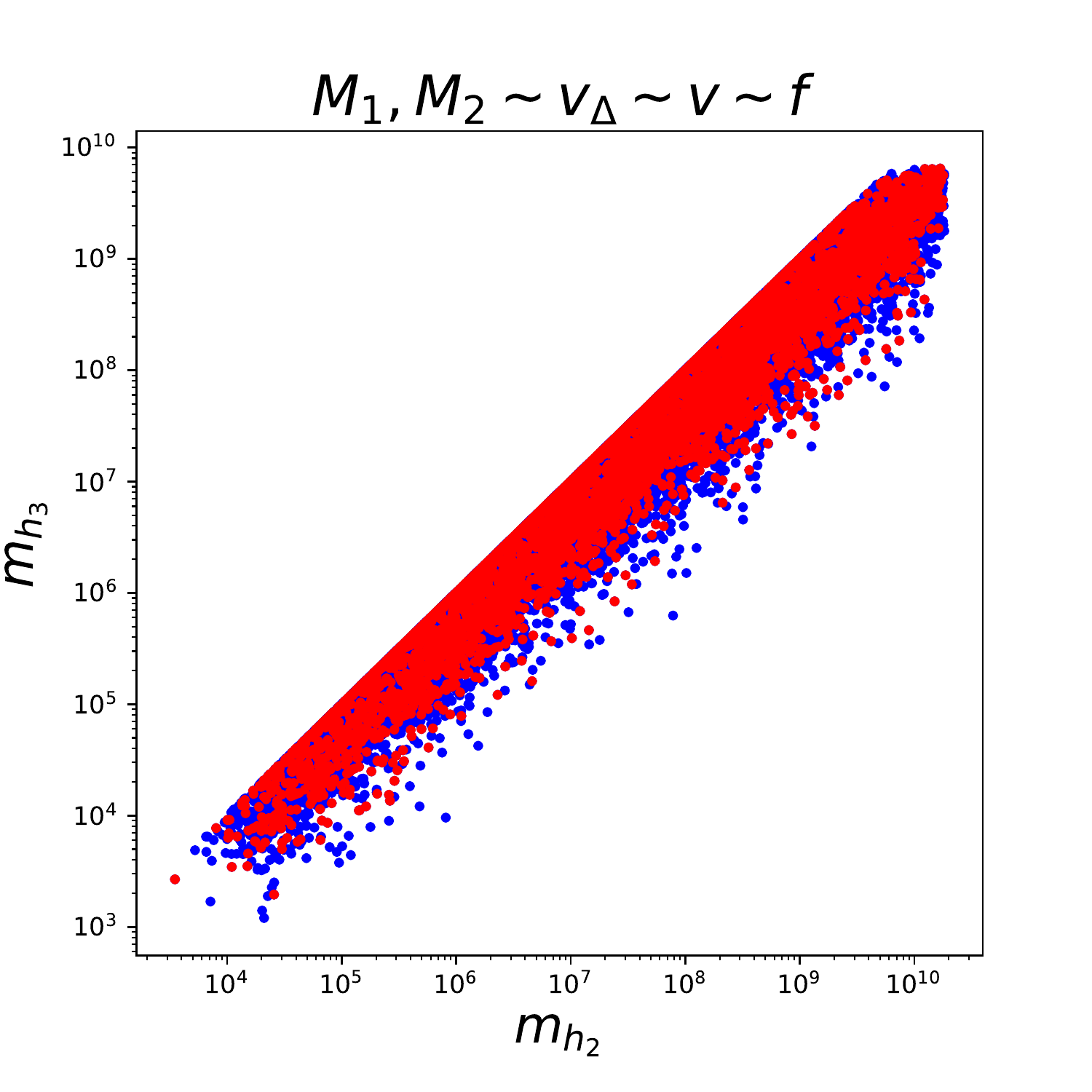} 
    \label{fig300b:a} 
    \vspace{4ex}
  \end{subfigure}
  \begin{subfigure}[b]{0.5\linewidth}
    \centering
\includegraphics[width=0.75\linewidth]{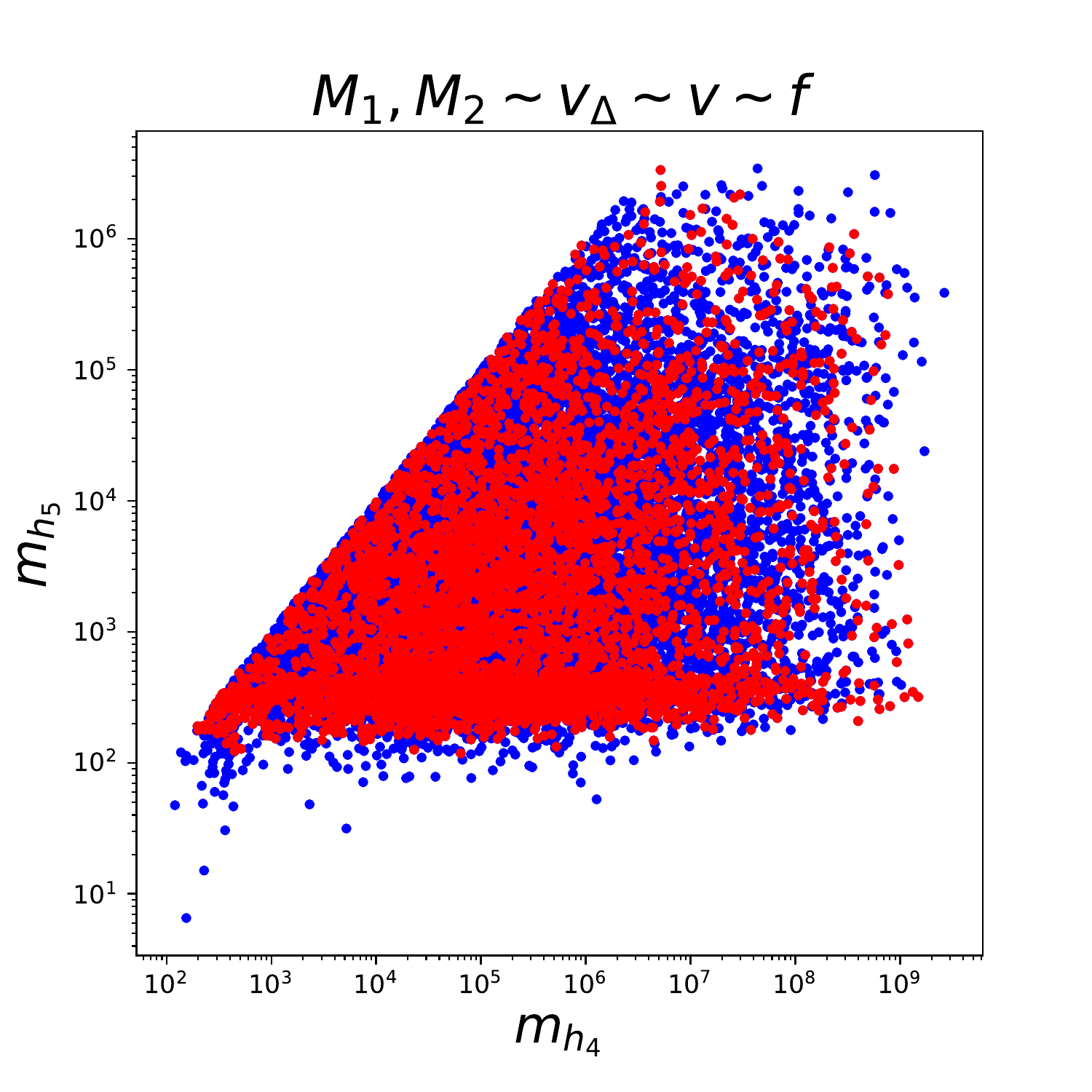} 
    \label{fig300b:b} 
    \vspace{4ex}
  \end{subfigure} 
  \begin{subfigure}[b]{0.5\linewidth}
    \centering    \includegraphics[width=0.75\linewidth]{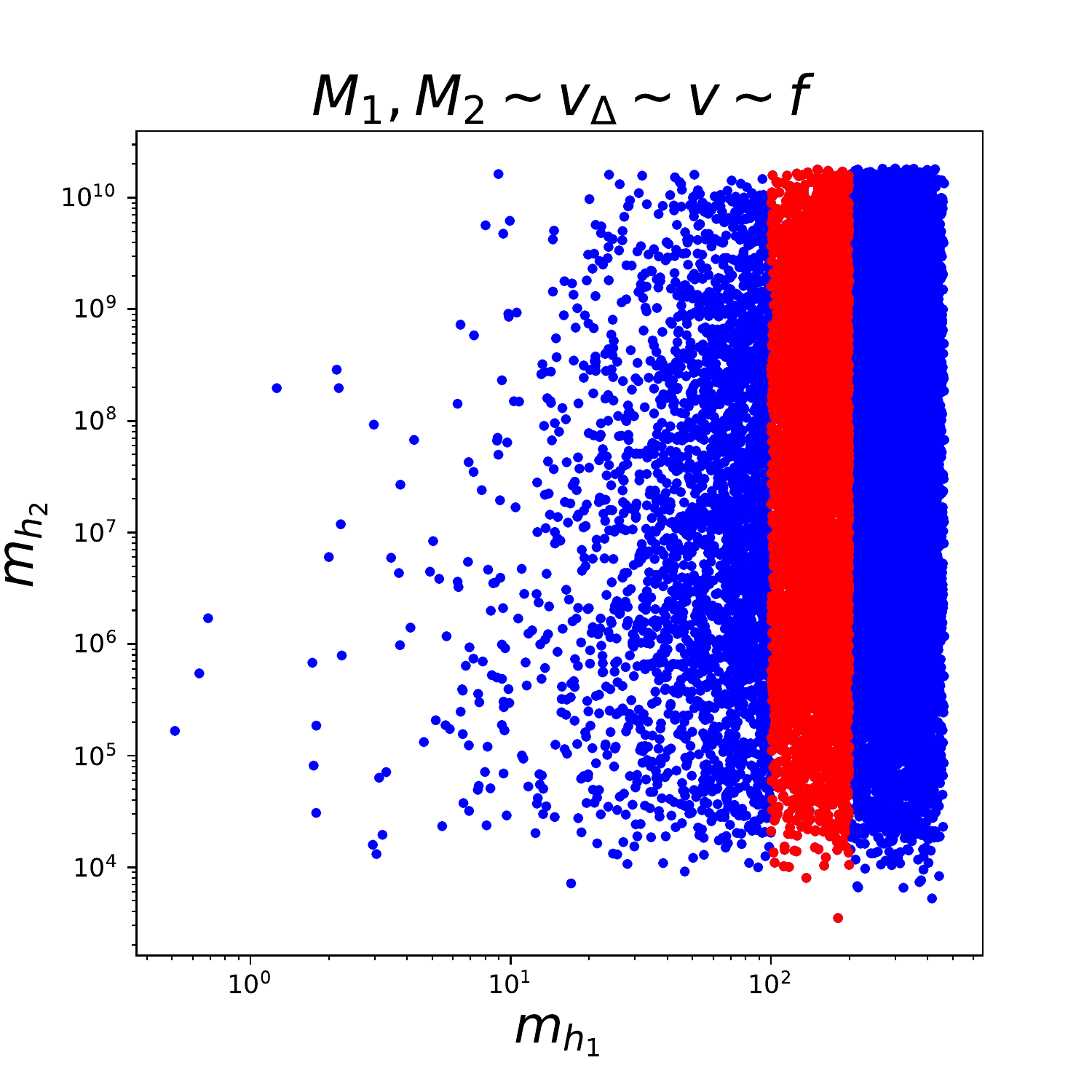}
    \label{fig300b:c} 
  \end{subfigure}
\caption{CP even at $ v_\Delta \sim M_1,M_2 \sim f\sim v$ . The blue dots represents all possible positive definite masses of the scalars at tree-level and the red dots represents the positive definite masses of the scalars with the Higgs-like particle mass ($m_{h_1}$) lying between $100$ GeV and $200$ GeV(See the text).}
  \label{fig300b} 
\end{figure}

\begin{figure}[ht] 
  \begin{subfigure}[b]{0.5\linewidth}
    \centering    \includegraphics[width=0.75\linewidth]{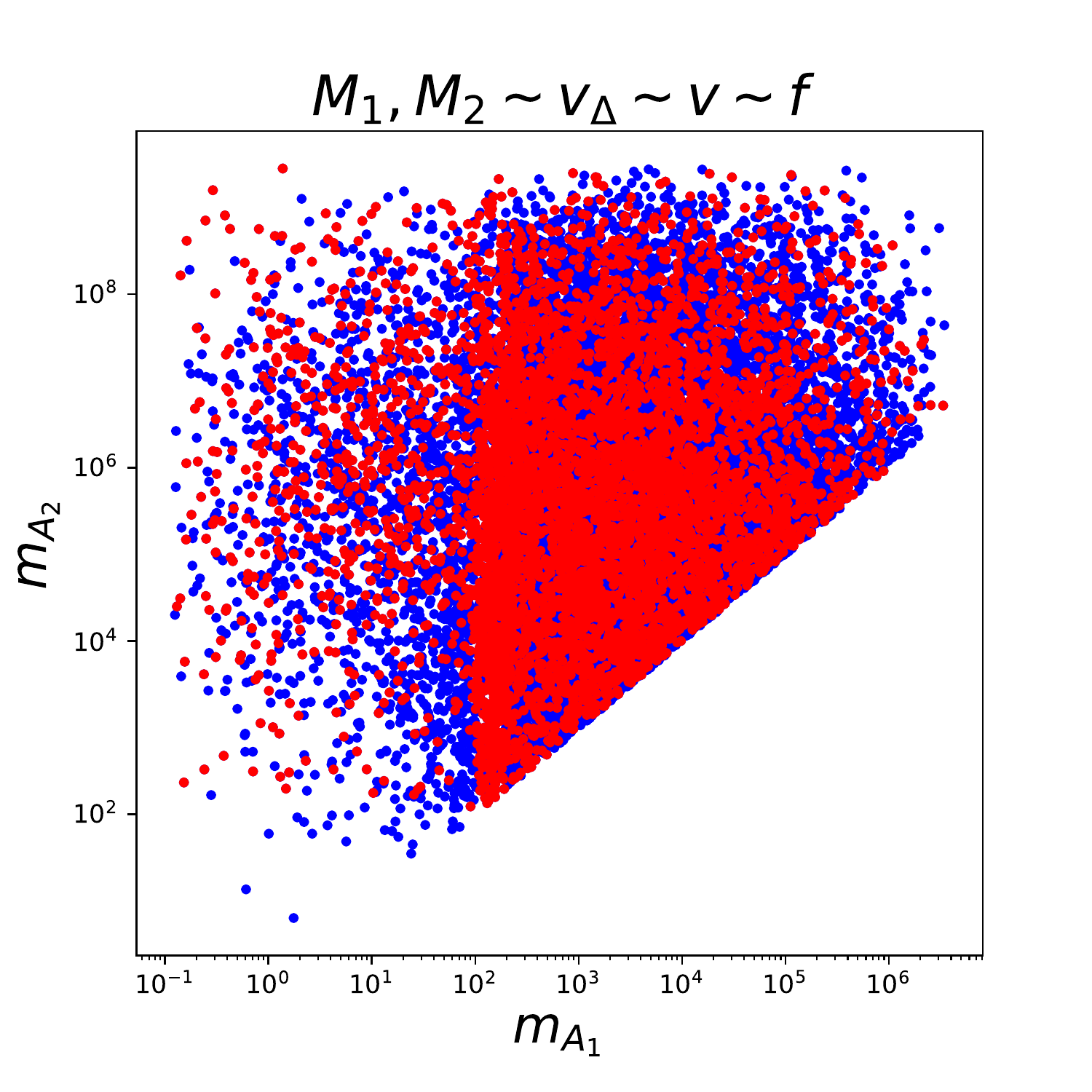}    
    \label{fig600b:a} 
    \vspace{4ex}
  \end{subfigure}
  \begin{subfigure}[b]{0.5\linewidth}
    \centering
\includegraphics[width=0.75\linewidth]{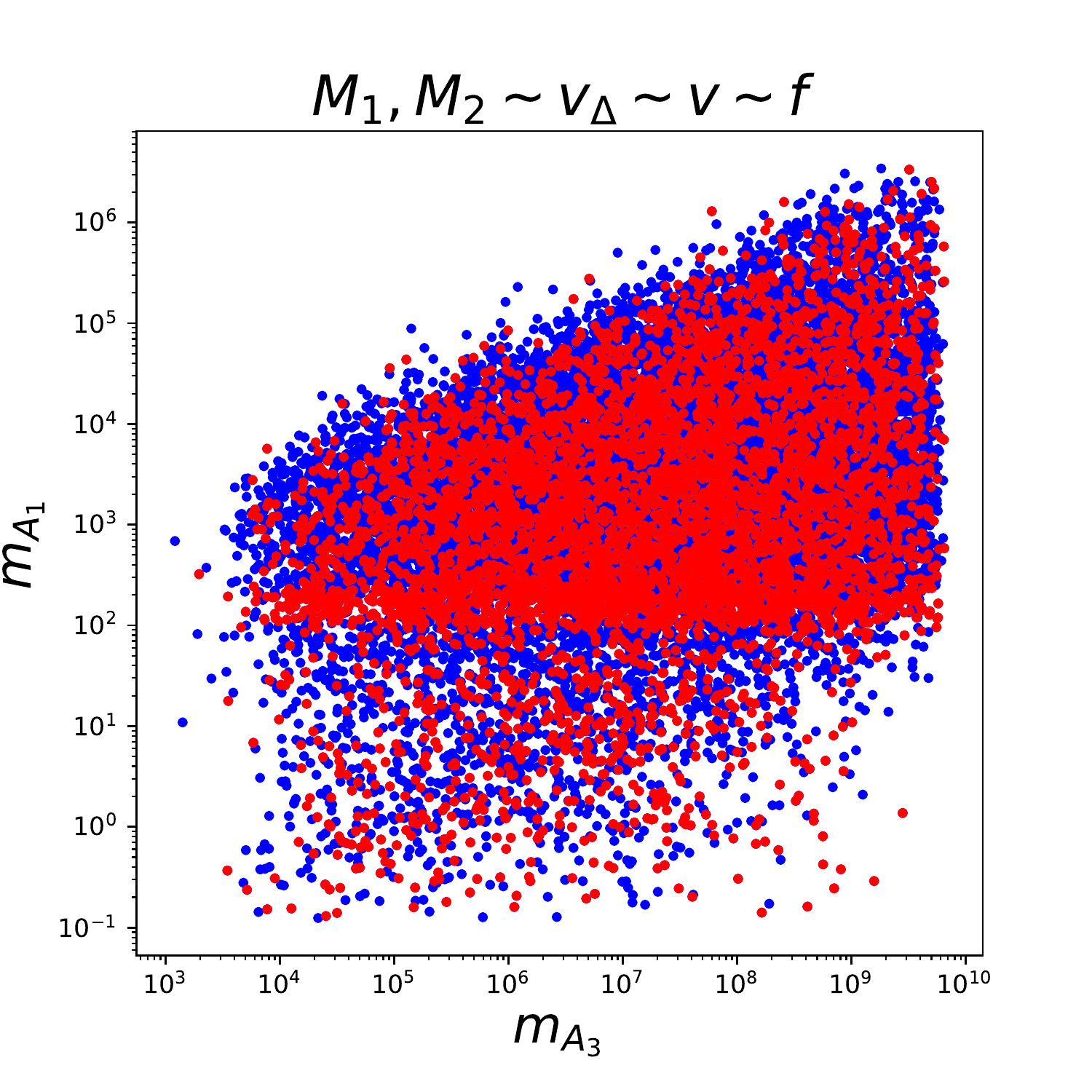}   
    \label{fig600b:b} 
    \vspace{4ex}
  \end{subfigure} 
 \caption{CP odd at $ v_\Delta \sim M_1,M_2 \sim f\sim v$. The blue dots represents all possible positive definite masses of the scalars at tree-level and the red dots represents the positive definite masses of the scalars with the Higgs-like particle mass ($m_{h_1}$) lying between $100$ GeV and $200$ GeV(See the text). }
  \label{fig600b} 
\end{figure}

\begin{figure}[ht] 
  \begin{subfigure}[b]{0.5\linewidth}
    \centering    \includegraphics[width=0.75\linewidth]{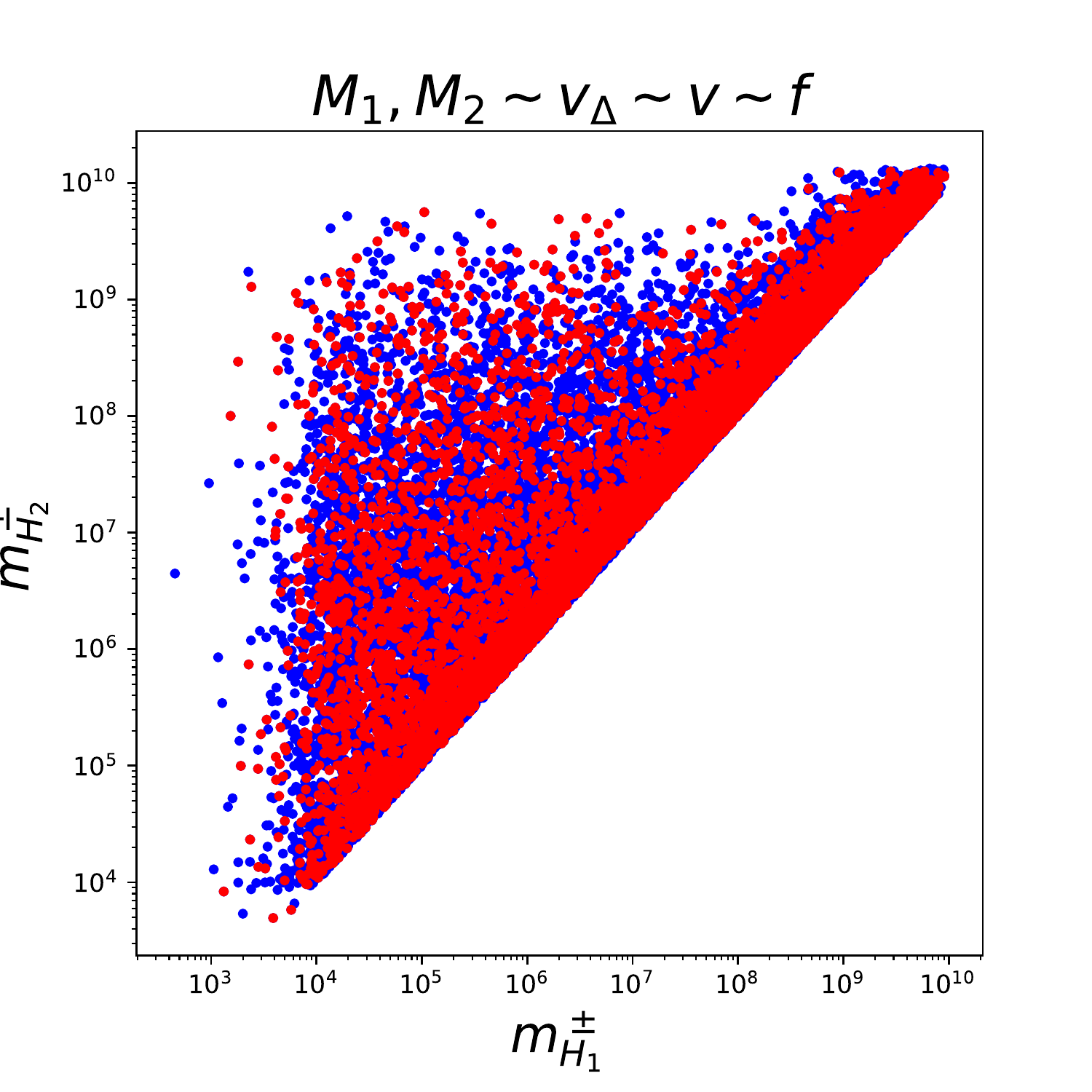}    
\caption{331 singly-charged scalars} 
    \label{fig900b:a} 
    \vspace{4ex}
  \end{subfigure}
  \begin{subfigure}[b]{0.5\linewidth}
    \centering
\includegraphics[width=0.75\linewidth]{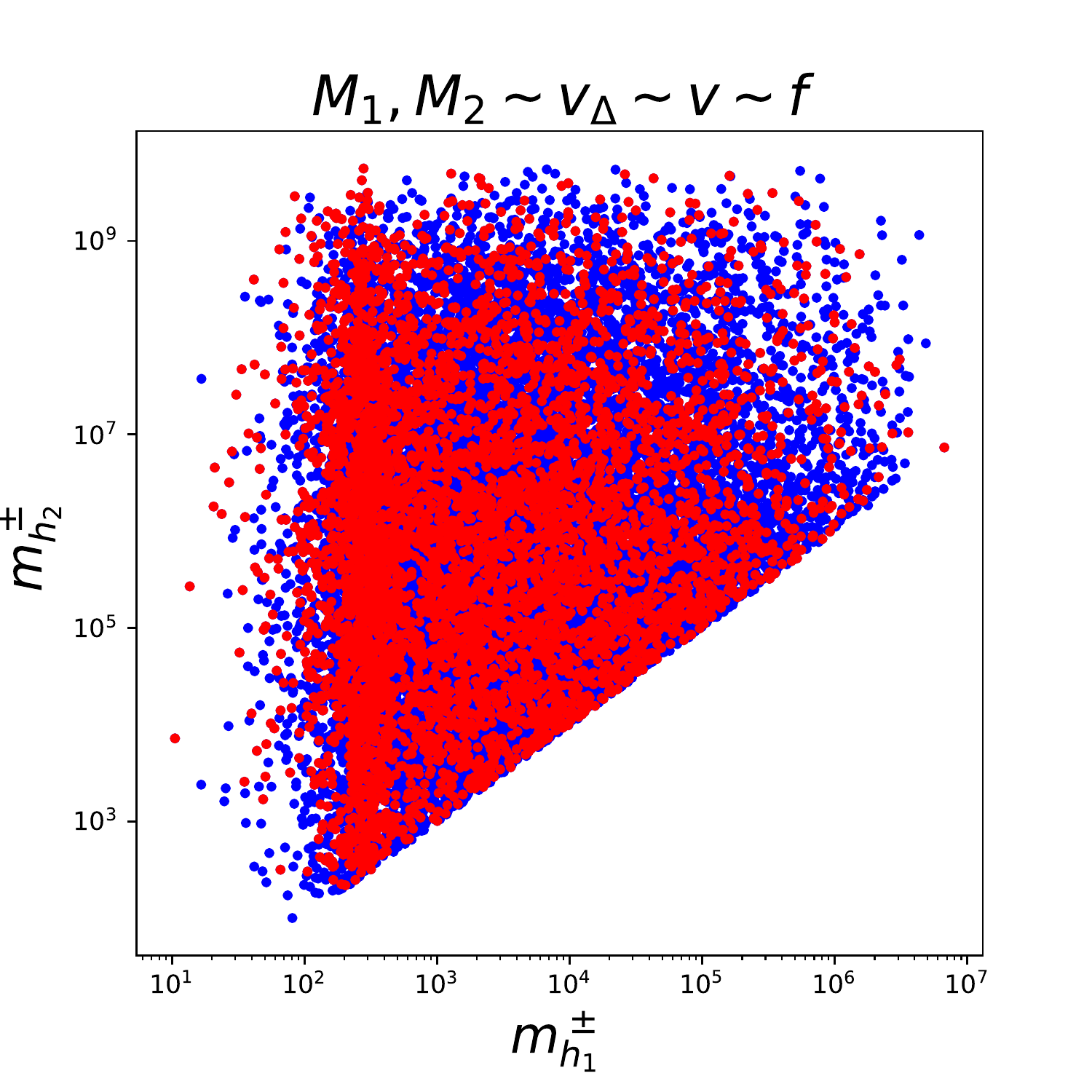}   
\caption{EW singly-charged scalars} 
    \label{fig900b:b} 
    \vspace{4ex}
  \end{subfigure} 
 \caption{Singly charged scalars at $ v_\Delta \sim M_1,M_2 \sim f\sim v$. The blue dots represents all possible positive definite masses of the scalars at tree-level and the red dots represents the positive definite masses of the scalars with the Higgs-like particle mass ($m_{h_1}$) lying between $100$ GeV and $200$ GeV(See the text).}
  \label{fig900b} 
\end{figure}

\begin{figure}[ht] 
    \centering    \includegraphics[width=0.4\linewidth]{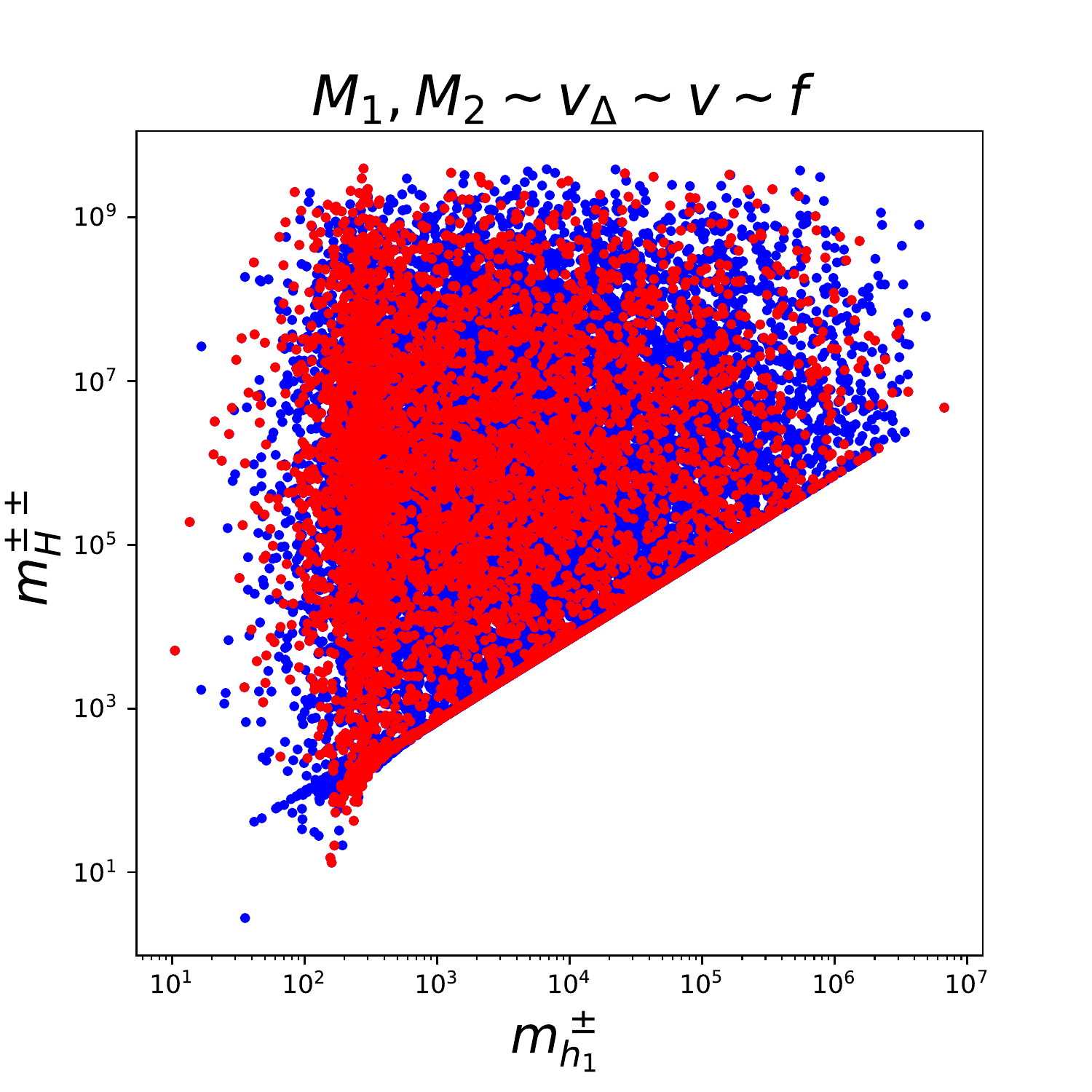}    \caption{Doubly charged scalar at $ v_\Delta \sim M_1,M_2 \sim f\sim v$. The blue dots represents all possible positive definite masses of the scalars at tree-level and the red dots represents the positive definite masses of the scalars with the Higgs-like particle mass ($m_{h_1}$) lying between $100$ GeV and $200$ GeV(See the text).}    
    \vspace{4ex}
  \label{fig1200b} 
\end{figure}

\begin{figure}[ht] 
    \centering    \includegraphics[width=0.4\linewidth]{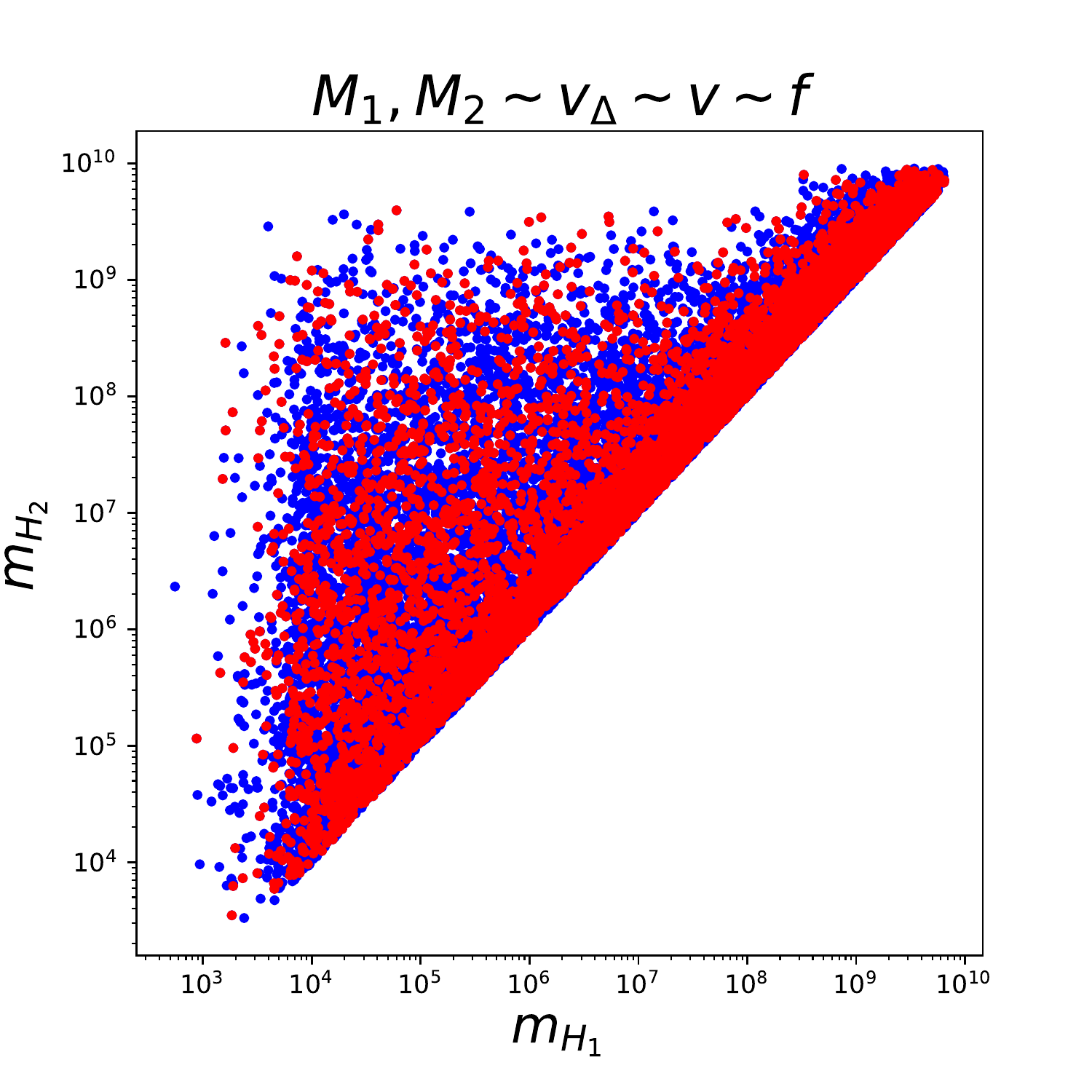}    \caption{Complex neutral scalar at $ v_\Delta \sim M_1,M_2 \sim f\sim v$. The blue dots represents all possible positive definite masses of the scalars at tree-level and the red dots represents the positive definite masses of the scalars with the Higgs-like particle mass ($m_{h_1}$) lying between $100$ GeV and $200$ GeV(See the text).}    
    \vspace{4ex}
  \label{fig1500b} 
\end{figure}

\clearpage

\subsection{Discussion of the results}
Here in this section we had exposed four different scenarios for the scalars of the theory. We had explored different energy regimes for the explicit and spontaneous lepton number symmetry breaking parameters $M_1$, $M_2$ and $v_\Delta$, respectively. However, before finishing this section, we would like to discuss about other analyses that we did for completeness and clarify why we choose specifically the four regimes described above. We had analysed, too, the following different cases: for $M_1$ and $M_2$ at very high energies ($10^8$ GeV$<M_1$,$M_2<10^{14}$ GeV), the vev $v_\Delta$ between ($10^{-12}$ GeV$<v_\Delta<10^{-4}$ GeV) and between ($10^{-4}$ GeV$<v_\Delta<1$ GeV). These two cases reveals the same phenomenology. The particles from the sextet totally decouples from the three triplet particles ($\chi$, $\eta$ and $\rho$) and all of its components becomes very heavy, in the order of $M_1$ and $M_2$ or higher.

\clearpage

\section{Conclusions}
In this work, we made a systematic study of the spectrum of scalars that  emerges from the scalar potential that engenders the type II seesaw mechanism of generation of small masses for the neutrinos inside the 331RHNs. The potential is composed by three triplets  and one sextet of scalars. Among the several terms that compose the potential there are two trilinear terms that breaks explicitly the lepton number and one that breaks explicitly the Peccei-Quinn symmetry. These terms are characterized by the energy  parameters $M_1$, $M_2$ and $f$, respectively. They determine the profile of the spectrum of scalars.

As remarkable case we refer to one in which $f$ belongs to the electroweak scale and $M_1$ and $M_2$ belong to sub-keV scale (obeying the hierarchy $v_\Delta \,, v_\sigma \ll M_1\,,\,M_2$ ). in it  the spectrum of scalars is obtained analytically. The nice result here is that the set of scalars with masses belonging to  the electroweak scale recovers the well know case THDM + triplet which has a rich phenomenology, while the other remaining scalars belongs to the 331 scale. 

For any other case concerning the parameters $M_1$, $M_2$, $v_\sigma$, $v_\Delta$ and $f$, the spectrum of scalars can be obtained only numerically. We did such study for a couple of cases. Here, we remark the case in which $M_1\,, M_2 \sim f \sim v$. This case can generate the 3HDM + triplet where, now, the third doublet belongs to the electroweak scale and it is dominantly composed by $\Phi$, the leptophilic doublet inside the sextet $S$.

In summary, in order to generate  small neutrino masses and engender the type II seesaw mechanism in the model, one needed to add a sextet of scalars to the original scalar content of the 331RHNs. Despite such extension generates a very  complex potential, fortunately it accommodates  a  phenomenological viable and attractive Higgs spectra for a wide range of the parameter space. Nonetheless, we can affirm that, for $f$, $M_1$ and $M_2$ varying up to a maximum of $v$ (electroweak scale) we  always get an effective THDM+triplet model separated from a sector with scalars belonging to the 331 typical energies. Such result is very deep, since, a very complex scalar sector can be associated directly with a well know phenomenology exhaustively studied in the literature and that it can be probed at present or future colliders. 

\clearpage
\section*{Acknowledgments}
C.A.S.P  was supported by the CNPq research grants No. 311936/2021-0 and J.P.P. has received funding/support from the European Union’s Horizon 2020 research and innovation program under the Marie Skłodowska-Curie grant agreement No
860881-HIDDeN.
\bibliography{bibliography}
\end{document}